\documentclass[preprint,12pt,authoryear]{elsarticle}

\usepackage[utf8]{inputenc}
\usepackage[T1]{fontenc}
\usepackage{amsthm, amsmath, amssymb,xcolor}
\usepackage{dcolumn}
\usepackage{braket}
\DeclareUnicodeCharacter{2009}{ }
\DeclareUnicodeCharacter{2013}{--}
\DeclareUnicodeCharacter{2212}{-}
\DeclareUnicodeCharacter{03BB}{$\lambda$}

\usepackage{hyperref}
\hypersetup{
    colorlinks=true,
    linkcolor=blue,
    filecolor=magenta,      
    urlcolor=cyan,
    }

\usepackage[symbol]{footmisc}
\usepackage{gensymb}
\usepackage{placeins}

\biboptions{numbers,sort&compress}
\newcommand*{\fig}[1]{Figure~\ref{fig:#1}}
\newcommand*{\tab}[1]{Table~\ref{tab:#1}}

\begin{document}
\begin{frontmatter}

\title{Understanding insulating ferromagnetism in $\text{LaCoO}_{3}$ films under tensile strain}

\author[a]{Ali Barooni}
\author[b]{Murod Mirzhalilov}
\author[b]{Mohit Randeria}
\author[c]{Patrick M. Woodward}
\author[a]{Maryam Ghazisaeidi\footnote{Corresponding Author}}

\affiliation[a]{organization={Department of Materials Science and Engineering, The Ohio State University},
            addressline={140 W 19th Ave}, 
            city={Columbus},
            postcode={43210}, 
            state={OH},
            country={USA}}
\affiliation[b]{organization={Department of Physics, The Ohio State University},
            addressline={191 W Woodruff Ave}, 
            city={Columbus},
            postcode={43210}, 
            state={OH},
            country={USA}}
\affiliation[c]{organization={Department of Chemistry and Biochemistry, The Ohio State University},
            addressline={151 W Woodruff Ave}, 
            city={Columbus},
            postcode={43210}, 
            state={OH},
            country={USA}}

\begin{abstract}

LaCoO$_3$ thin films grown under epitaxial tensile strain exhibit a robust ferromagnetic insulating state that is absent in the bulk. Despite many studies, both experimental and computational, the microscopic origin of this phenomenon is not well understood. In this work, density functional theory calculations are used to systematically investigate the magnetic ground state of stoichiometric  LaCoO$_3$ under epitaxial strain equivalent to that imposed by a SrTiO$_3$ substrate. The results identify a ferromagnetic insulating ground state characterized by a unique ordered array of high-spin (HS) and low-spin (LS) Co$^{3+}$ ions. The spin state ordering is best described as 2 $\times $ 2 columns that consist of alternating HS and LS Co$^{3+}$ ions, separated by planes of LS Co$^{3+}$ ions. This leads to HS–LS–LS repeating sequence of Co$^{3+}$ ions in both pseudocubic [100] and [010] directions. Analysis of the electronic structure confirms the presence of an insulating gap. Evaluation of the superexchange interactions reveal ferromagnetic interactions between HS Co$^{3+}$ ions via 90$^\circ$ paths, and antiferromagnetic interactions via 180$^\circ$ paths, both of which are facilitated by empty $\sigma^*$ ($e_g$) orbitals on the diamagnetic LS Co$^{3+}$ ions. The strength and number of 90$^\circ$ ferromagnetic interactions are sufficient to overcome the competing 180$^\circ$ antiferromagnetic interactions stabilizing a ferromagnetic insulating state.



\footnotetext{Email addresses: barooni.1@osu.edu(Ali Barooni), ghazisaeidi.1@osu.edu(Maryam Ghazisaeidi)}

\end{abstract}

\begin{keyword}
First-principles calculations \sep Transition-metal oxides \sep Ferromagnetic insulator \sep Epitaxial strain \sep LaCoO$_3$ thin films \sep Spin-state transition


\end{keyword}

\end{frontmatter}

\section{Introduction}
\label{sect:intro}

The interplay between magnetism and electronic transport in transition metal oxides (TMOs) is not only a rich source of complex fundamental materials science, but also underpins a wide range of applications from spintronics to thermoelectric energy harvesting~\cite{morosan_strongly_2012}. The combination of ferromagnetism and insulating behavior is of particular interest, but this combination is rarely seen in TMOs, where ferromagnetism is typically stabilized by the double-exchange mechanism, which requires delocalized electrons and leads to metallic conductivity \cite{kanamori_superexchange_1959}. Stabilization of an insulating ferromagnetic state requires a subtle interplay of charge and/or orbital ordering, superexchange interactions, and lattice degrees of freedom\cite{tokura_orbital_2000,tokura_correlated-electron_2003}. Despite their rarity, ferromagnetic insulators are of great interest because of their ability to facilitate the transport of pure spin currents with minimal charge flow, a prerequisite for next-generation spintronic technologies, including quantum information processing and energy-efficient devices \cite{hellman_interface-induced_2017,tannous_magnetic_2017,wang_room-temperature_2017,deng_gate-tunable_2018}.

In particular, Lanthanum cobalt oxide ($\text{LaCoO}_3$)  has attracted significant attention due to its unusual magnetic properties. In its bulk form, LaCoO$_3$ is a diamagnetic insulator at low temperatures, with all cobalt ions ($\text{Co}^{3+}$) in the low-spin (LS) state ($t_{2g}^6$$e_{g}^0$, $S=0$)\cite{asai_neutron-scattering_1994}. Transitions from the LS state to either the intermediate-spin (IS) ($t_{2g}^5$$e_{g}^1$, $S=1$) or high-spin (HS) ($t_{2g}^4$$e_{g}^2$, $S=2$) state can be driven by changes in temperature~\cite{asai_temperature-induced_1989}, defects~\cite{hansteen_crystal_1998}, or external pressure~\cite{panfilov_pressure_2018}. 
These transitions are intimately linked to changes in the crystal structure and the volume of the $\text{CoO}_6$ octahedra, which modify the balance between Hund’s exchange energy ($\Delta_{EX}$) and the crystal-field splitting ($\Delta_{CF}$) that governs the Co spin states.

Interestingly, when grown as thin films on substrates that induce tensile strain, such as  $\left( \text{LaAlO}_3 \right)_{0.3} \left( \text{Sr}_2\text{TaAlO}_6 \right)_{0.7}$ (LSAT) or $\text{SrTiO}_3$ (STO), LaCoO$_3$ exhibits ferromagnetic insulating behavior with Curie temperatures $\text{T}_\text{C}$ in the range of 65–80 K \cite{li_strain_2018, liu_two-dimensional_2021,shin_tunable_2022}. In contrast, films grown on substrates that impose compressive strain, such as LaAlO$_3$, remain nonmagnetic or weakly paramagnetic down to low temperatures \cite{choi_strain-induced_2012}. It is generally assumed that tensile strain modifies the distribution of spin states at low temperature, but a satisfactory explanation of how that translates into an insulating ferromagnetic ground state remains elusive. In addition to epitaxial strain, oxygen vacancies have frequently been invoked as a contributing factor, since oxygen deficiency can stabilize Co$^{2+}$ ions and promote ferromagnetic coupling through mixed-valence interactions. This has led to interpretations in which the observed magnetism arises from oxygen-vacancy–driven phases or from an interplay between strain and non-stoichiometry. However, ferromagnetic insulating behavior has been reported in films, where electron energy loss spectroscopy detected no evidence of Co$^{2+}$ ions, thereby ruling out the presence of oxygen vacancies \cite{choi_strain-induced_2012,yoon_strain-induced_2021}. Thus it would seem that epitaxial strain alone is sufficient to stabilize a ferromagnetic insulating phase.

Several fundamental issues regarding the origins of ferromagnetism in LaCoO$_3$ films remain unresolved. How does the imposition of tensile strain affect the structure, in particular the Co–O bond distances that are intimately tied to the spin state? What mixture of spin-states (low, intermediate, or high-spin) is present and what is the ordered pattern of those spin states. Studies that invoke rocksalt-type ordering of HS and LS Co$^{3+}$ ions \cite{seo_strain-driven_2012, chen_spin_2023,li_emergent_2023} are difficult to reconcile with conventional superexchange considerations \cite{anderson_new_1959}, which generally predict antiferromagnetic coupling through nearly linear Co(HS)–O–Co(LS)–O–Co(HS) bonds. 

Given the difficulties associated with interrogating the structures of thin films at low temperatures, first-principles studies can offer important insights into the origins of ferromagnetism in LaCoO$_3$ films. However, existing first-principles studies still pose important limitations \cite{rondinelli_structural_2009,seo_strain-driven_2012,meng_strain-induced_2018,geisler_competition_2020,yoon_strain-induced_2021}. Many employ simulation cells that are too small to fully capture the complexities of octahedral tilting and exclude potentially viable patterns of spin-state ordering. Moreover, the common practice of applying a uniform Hubbard $U_\text{eff}$ to all Co sites may obscure important site-dependent variations in electronic structure and spin state. These constraints have hindered a clear identification of the magnetic ground state and the dominant exchange mechanisms.

In this work, we focus exclusively on the role of epitaxial strain and investigate the magnetic ground state of stoichiometric LaCoO$_3$ using density functional theory calculations, with sufficiently large supercells, under tensile strain equivalent to that imposed by an STO substrate. We first determine the strain-stabilized magnetic ground state and then analyze its electronic structure. Next, we extract the magnetic coupling parameters to quantitatively characterize the magnetic interactions. Finally, we examine the superexchange pathways responsible for the magnetic ordering and provide a theoretical explanation for the observed behavior.

\section{Computational Methods}
\label{sect:methods}

Density functional theory (DFT) calculations were performed using the Vienna Ab initio Simulation Package (VASP)\cite{kresse_efficiency_1996,kresse_efficient_1996}. The projector augmented-wave (PAW) method was employed. Structural relaxations were carried out using the meta–generalized gradient approximation (meta-GGA) exchange–correlation functional in the regularized-restored strongly constrained and appropriately normed ($r^2$SCAN) formulation\cite{furness_accurate_2020}. 
For density of states (DOS) calculations, on-site electron correlations were treated within the DFT+$U$ formalism ($r^2$SCAN+$U$) using the Dudarev approach\cite{dudarev_electron-energy-loss_1998}, with an effective Hubbard parameter $U_{\mathrm{eff}} = U - J = 1$~eV applied to the Co $3d$ states. 
A plane-wave kinetic energy cutoff of 550~eV was used throughout all calculations.
For bulk LaCoO$_3$, the Brillouin zone was sampled using a $6\times6\times6$ Monkhorst–Pack $k$-point mesh. 
Supercell calculations were performed using a $1\times1\times4$ $\Gamma$-centered $k$-point mesh. 
Structural relaxations were continued until the total energy converged to within $10^{-6}$~eV and the Hellmann–Feynman forces on each atom were smaller than 5~meV/\AA.
For the calculation of magnetic exchange coupling parameters, a stricter electronic convergence criterion of $10^{-8}$~eV was employed, and a $2\times2\times2$ $k$-point mesh was used for all magnetic configurations. The valence electron configurations used in the PAW potentials are as follows: La: $5p^6 5d^1 6s^2$; Co: $3d^7 4s^2$; and O: $2s^2 2p^4$.

\section{Results}

\begin{figure}[!h]
    \centering
    \includegraphics[width=1\textwidth]{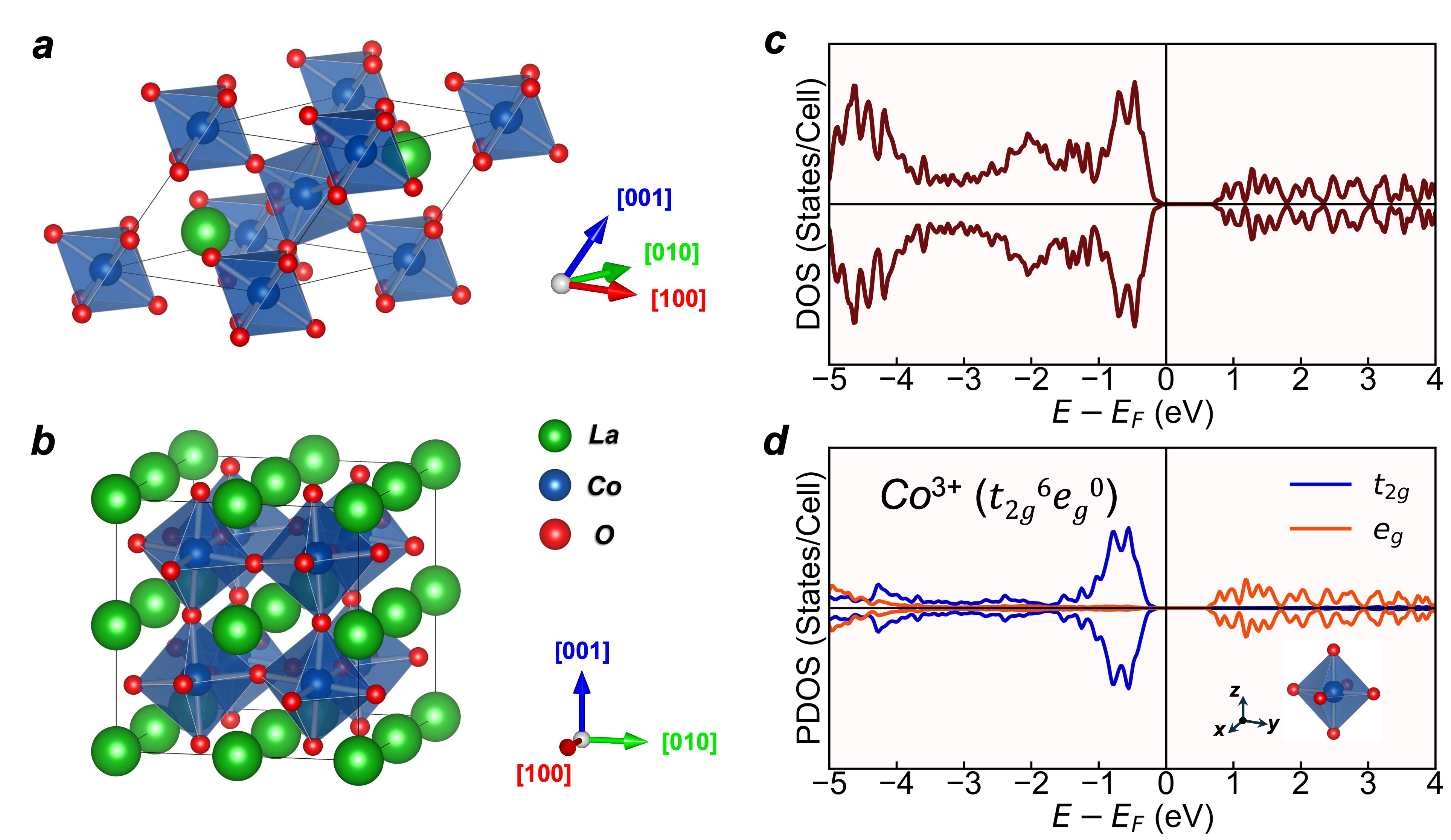}
    \caption{
    Crystal structure and electronic density of states (DOS) of bulk LaCoO$_3$. (a) and (b) show the $R\bar{3}c$ crystal structure and its corresponding pseudocubic unit cell with octahedral tilting respectively.  The DOS of bulk LaCoO$_3$ reveals a band gap of 0.64 eV in (c). (d) shows the partial density of states (PDOS) of $\text{Co}^{3+}$. Integration of the PDOS confirms fully occupied $t_{2g}$ orbitals, consistent with a LS state.
    }
    \label{fig:LCOcrystal}
\end{figure}

Bulk LaCoO$_3$ has a rhombohedral structure with space group $R\bar{3}c$ at room temperature~\cite{kobayashi_structural_2000}, as shown in \fig{LCOcrystal}(a). As a reference, the bulk properties of LaCoO$_3$ are first calculated. The fully relaxed lattice parameters are summarized in \tab{latticeparams}. \fig{LCOcrystal}(e) and \fig{LCOcrystal}(f) show the calculated total DOS and the projected DOS (PDOS) for the Co ions, respectively. The results indicate an insulating ground state with a band gap of approximately 0.64 eV, in good agreement with previously reported values~\cite{chainani_electron-spectroscopy_1992}.
Integration of the Co-projected DOS reveals that all Co ions adopt an LS configuration, characterized by fully occupied $t_{2g}$ orbitals and empty $e_g$ states. This result is consistent with previous experimental studies of bulk LaCoO$_3$ at low temperatures\cite{asai_neutron-scattering_1994}.
\fig{LCOcrystal}(b) illustrates the corresponding pseudocubic bulk unit cell containing eight Co ions and incorporating the characteristic $a^{-}a^{-}a^{-}$ octahedral tilting pattern. We introduce this pseudocubic 
representation because it matches the crystallographic orientation of LaCoO$_3$ films grown on the STO substrate. This unit cell will therefore serve as the structural reference for the strained-bulk calculations discussed in the following section.

\begin{table}[ht]
    \centering
    
    \begin{tabular}{cccccccc}
    \hline 
        Method &  
        $a$ (\AA) & 
        $\beta \ (^\circ)$ & 
        $r_{Co-O}$ (\AA) & 
        $\theta_{Co-O-Co} \ (^\circ)$ & 
        $a_{pc}$ (\AA) \\
    \hline
        DFT (This work) &
        5.3594 &
        60.970 &
        1.929 &
        163.23 &
        3.817 \\
        
        Exp\cite{bull_low-temperature_2016}
        @ 4.2 K&
        5.34058 &
        60.988 &
        1.923 &
        163.1 &
        3.8046 \\
        
        Exp\cite{xu_structural_2001}
        @ 100 K &
        5.3781 &
        60.81 &
        1.922 &
        164.1 &
        3.8029 \\
        
    \hline
    \end{tabular}
    \caption{Lattice parameters, bond lengths, bond angles, and pseudocubic lattice constant of bulk LaCoO$_3$. Current DFT calculations agree well with experimental values.}

    \label{tab:latticeparams}
\end{table}

To assess the effect of epitaxial strain on the spin-state transition of a single CoO$_6$ octahedron, an in-plane biaxial tensile strain ranging from 0 to 4.0\% was applied, and the total energies of both spin configurations were compared. Above a critical tensile strain of approximately 1.8\%, the HS configuration becomes energetically favorable compared to the LS state (see Supporting Information). The results demonstrate that sufficiently large tensile strain alone can drive a spin-state transition of a single CoO$_6$ octahedron. Moreover, in these calculations, the IS state was not considered, as biaxial strain within this range did not stabilize an IS configuration.

\begin{figure}[!h]
    \centering
    \includegraphics[width=1\textwidth]{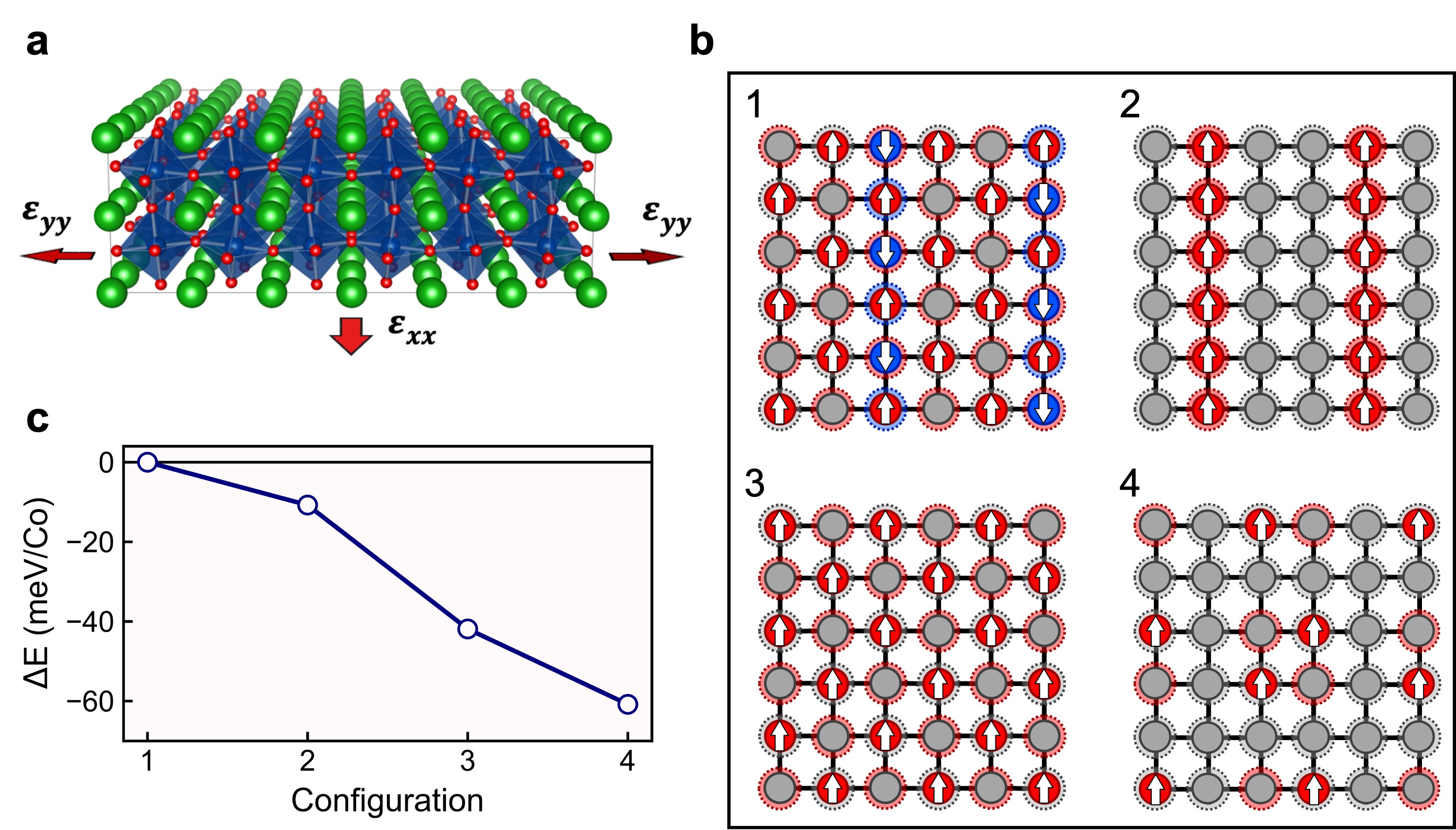}
    \caption{Selected magnetic configurations of the 3$\times$3$\times$1 supercell of pseudocubic LaCoO$_3$. (a) An in-plane epitaxial strain of 2.4\% was applied, with the out-of-plane lattice constant allowed to relax for all initial magnetic configurations. (b) The initial magnetic configurations of the previously presented models\cite{yoon_strain-induced_2021, kwon_nanoscale_2014,seo_strain-driven_2012}, mapped onto our supercell size, along with the lowest-energy configuration from our calculations (configuration 4), are shown. The red, blue, and gray colors represent HS ($\uparrow$), HS ($\downarrow$), and LS Co ions, respectively. The ionic arrangement in the top layer is indicated by solid lines, and the bottom layer is shown using dashed lines. (c) The energy differences among the magnetic configurations indicate that, under epitaxial tensile strain, our magnetic configuration is lower in energy than the previously proposed models.
    }
    \label{fig:LCOsupercell}
\end{figure}

Based on this finding, a $3 \times 3 \times 1$ supercell was constructed from the pseudocubic unit cell, as shown in \fig{LCOsupercell}(a). This supercell size provides sufficient degrees of freedom to accommodate different spin configurations and magnetic interactions while keeping the computational cost manageable. In particular, a larger in-plane cell allows a finite concentration of high-spin Co ions to interact without becoming unrealistically dilute. Strained LaCoO$_3$ was then modeled under epitaxial conditions corresponding to an STO substrate by fixing the in-plane lattice constants along the [100] and [010] directions to the STO lattice parameter ($a = 3.905$ \AA), which corresponds to a tensile strain of approximately 2.4\% relative to bulk LaCoO$_3$. The out-of-plane lattice constant was fully relaxed to accommodate the imposed biaxial strain.

This supercell allows for a broad range of magnetic configurations of the $\text{Co}^{3+}$ ions. For each of the 36 Co ions in the first Co layer of the supercell, three possible initial spin states were considered: HS ($\uparrow$), HS ($\downarrow$), and LS states. To reduce the configuration space, only magnetic arrangements preserving symmetry along both the [100] and [010] in-plane directions were retained, as there is no physical justification for preferential spin-state ordering along a single in-plane axis. In addition, configurations inconsistent with established magnetic exchange interaction rules\cite{goodenough_interpretation_1958} were not included in the primary set of configurations considered here.
Upon applying periodic boundary conditions in the in-plane directions, symmetry-equivalent configurations were eliminated. Considering the presence of the second Co layer along the out-of-plane direction of the supercell further reduced the number of physically distinct magnetic configurations to 95 (See supporting information).

Following full structural relaxations, a ferromagnetic ground state was identified, characterized by $2 \times 2$ rocksalt-type HS/LS regions separated by planes of LS Co ions (\fig{LCOsupercell}(b)). The relative energies of the various configurations are presented in \fig{LCOsupercell}(c), where the ground-state structure is energetically favored by at least 18.1~meV compared to other configurations reported in the literature\cite{yoon_strain-induced_2021, kwon_nanoscale_2014,seo_strain-driven_2012} which were also evaluated for comparison. Energy comparisons for additional configurations are provided in the Supporting Information. This ground state configuration exhibits an HS fraction of 22.2\% (16 out of 72 Co ions), corresponding to a net magnetization of 0.88 $\mu_B$ per Co ion, and an out-of-plane lattice constant of $c = 3.77$ \AA, in agreement with previously reported experimental values (c $\approx$ 3.76--3.79 \AA)\cite{fan_significant_2023,russell_understanding_2026}.

\begin{figure}[!h]
    \centering
    \includegraphics[width=1\textwidth]{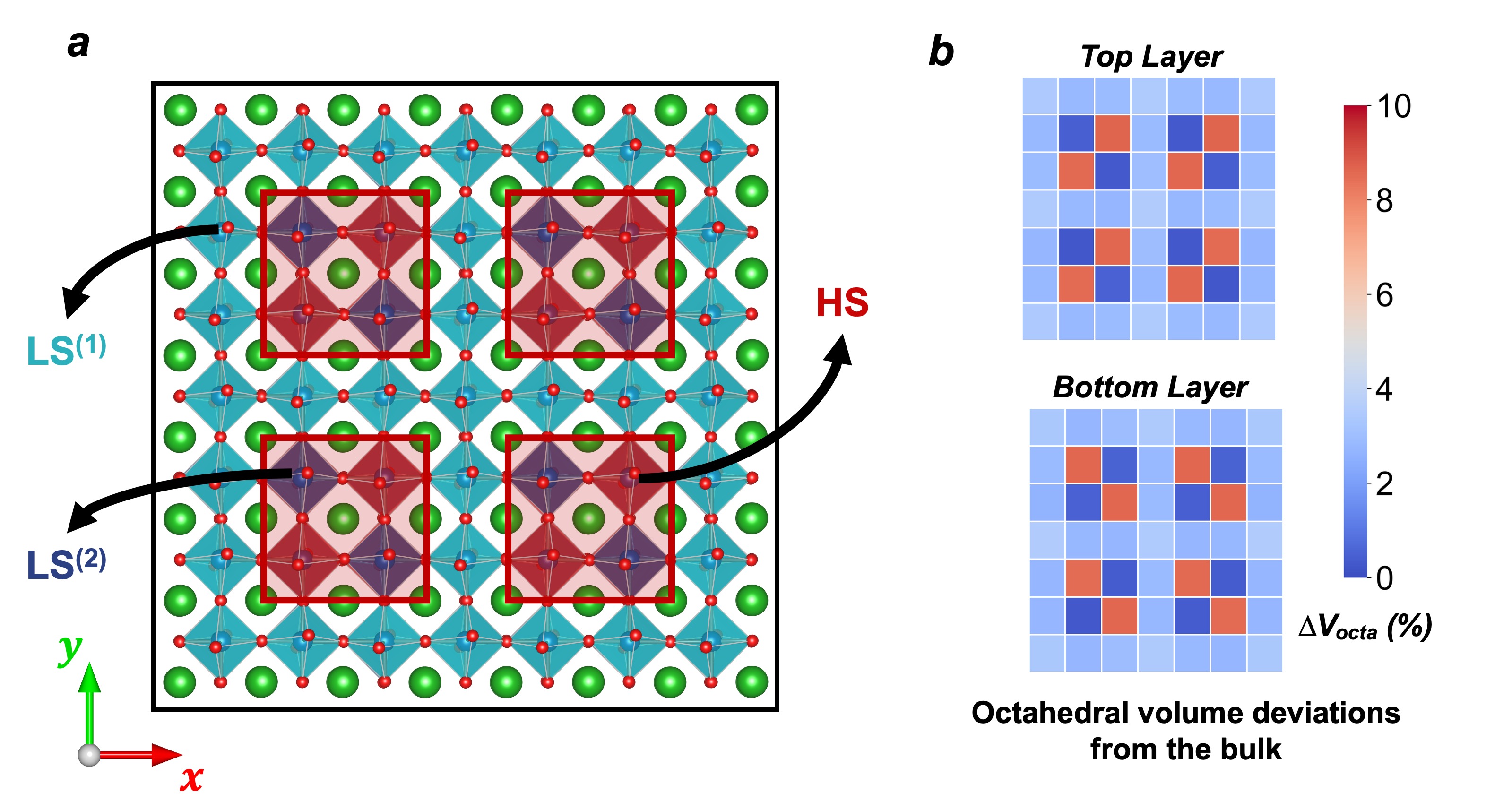}
    \caption{Magnetic structure and octahedral volumes of the ferromagnetic columnar model. (a) Ferromagnetic domains, composed of $2 \times 2$ rocksalt-type HS/LS $\text{Co}^{3+}$ ions, are separated by planes of LS ions. $\textit{LS}^{(1)}$ refers to the LS ions located in the separating LS planes (cyan), while $\textit{LS}^{(2)}$ denotes the LS ions within the columns (dark blue). (b) Deviations of the octahedral volumes from bulk values are shown for the top and bottom layers. The HS octahedra exhibit a larger volume increase compared to the LS octahedra.
    }
    \label{fig:FMC}
\end{figure}

The ground state configuration was further analyzed to examine the local structural distortions, specifically bond distances and bond angles, and to confirm its insulating character. \fig{FMC}(a) presents the top view of the ground state model, where HS $\text{CoO}_6$ octahedra are highlighted in red and LS octahedra in blue. As shown, each 2$\times$2 HS/LS region (indicated by red squares), referred to here as a ferromagnetic column, is separated by planes of LS $\text{Co}^{3+}$ ions. We refer to this ground state configuration as the \textit{ferromagnetic columnar model}. A closer inspection of the spin pattern reveals that two LS ions are always positioned between adjacent HS ions belonging to neighboring ferromagnetic columns in both the x and y directions. In this model, two distinct types of LS $\text{Co}^{3+}$ ions emerge, differentiated by their local environments and octahedral volumes: $\textit{LS}^{(1)}$ ions, positioned within the separating LS planes, and $\textit{LS}^{(2)}$ ions, located within the rocksalt-type HS/LS columns.

\fig{FMC}(b) presents the octahedral volume deviation from the bulk ($\text{V}_{b}=9.57$ \AA$^3$) within the top and bottom layers of the film. The HS octahedra exhibit an approximate 8.5\% increase in volume compared to the bulk, while the $\textit{LS}^{(1)}$ and $\textit{LS}^{(2)}$ octahedra show a more modest expansion of 3.0\% and 0.4\%, respectively. The average Co–O bond lengths and Co–O–Co bond angles for HS, $\textit{LS}^{(1)}$, and $\textit{LS}^{(2)}$ $\text{Co}^{3+}$ ions are summarized in \tab{bonds}. 
The HS CoO$_6$ octahedra exhibit an in-plane Co-O bond elongation of approximately 4.1\%, whereas the in-plane bond lengths of the two types of LS ($\textit{LS}^{(1)}$ and $\textit{LS}^{(2)}$) octahedra increase by about 1.5\% and remain nearly unchanged, respectively. Correspondingly, the octahedral volumes and in-plane Co-O bond lengths of HS Co$^{3+}$ ions are noticeably larger than those of LS ions, consistent with the occupancy of half-filled $e_g$ orbitals inherent to the HS configuration. The in-plane Co-O-Co bond angles change only slightly relative to the bulk structure, and the HS-O-LS$^{(1)}$ and HS-O-LS$^{(2)}$ angles are essentially identical; therefore, the spin state configuration is primarily determined by bond lengths rather than bond angles.


\begin{table}[ht]
    \centering
    \begin{tabular}{c}
        \includegraphics[width=0.5\textwidth]{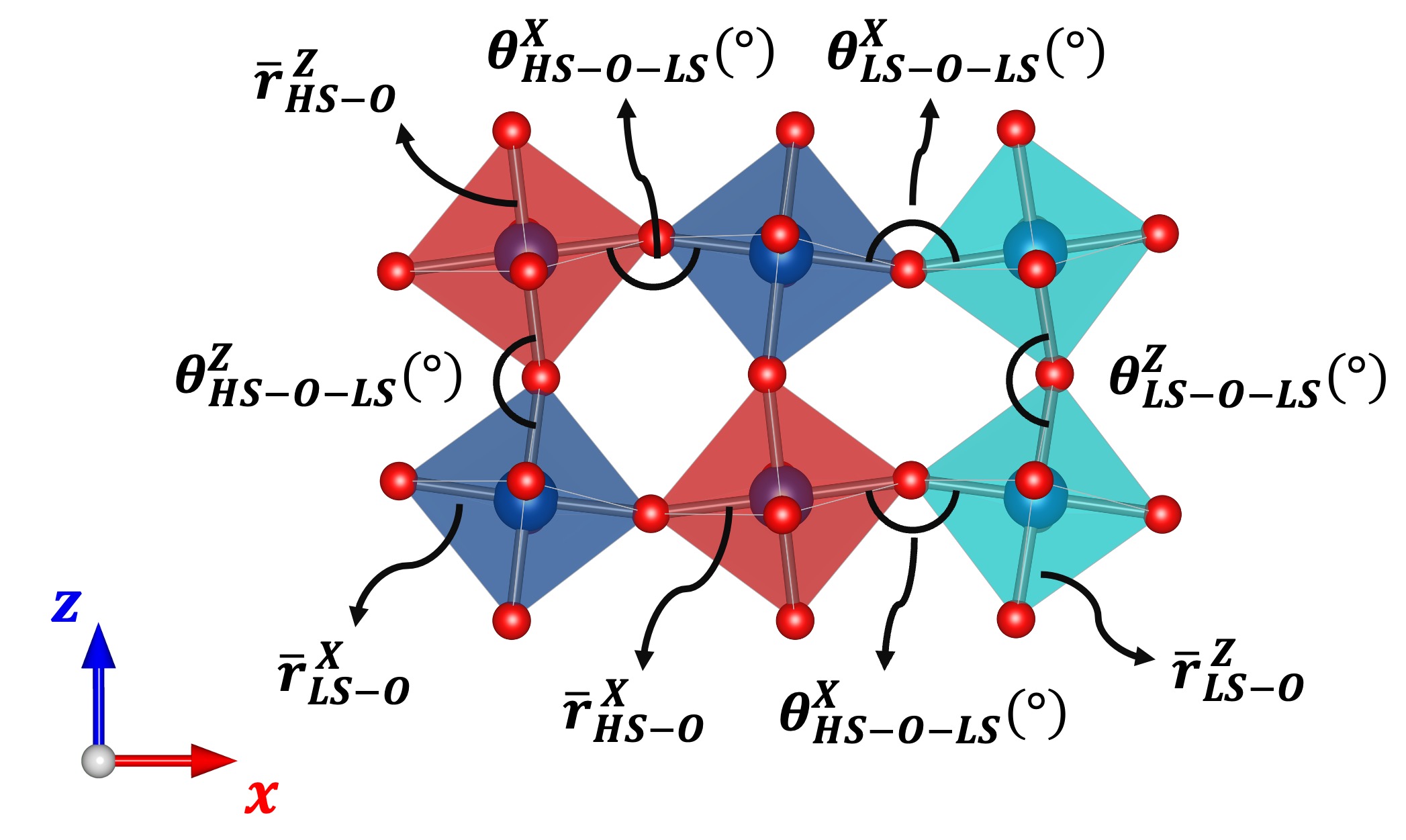} \\
    \end{tabular}

    \vspace{1em}

    \begin{tabular}{c | c c c c}
    \centering
        {Spin State} &
        $\bar{V}_{\text{octa}}$ (\AA$^3$) &
        $\bar{r}_{\text{Co--O}}^X$ (\AA) &
        $\bar{r}_{\text{Co--O}}^Y$ (\AA) &
        $\bar{r}_{\text{Co--O}}^Z$ (\AA) \\
    \hline
        Bulk (\textit{LS})& 9.57 & 1.93 & 1.93 & 1.93 \\
        $LS^{(1)}$ & 9.86 & 1.96 & 1.96 & 1.92 \\
        $LS^{(2)}$ & 9.61 & 1.93 & 1.93 & 1.90 \\
        \textit{HS} & 10.39 & 2.01 & 2.01 & 1.92 \\
    \end{tabular}
    
    \vspace{1em}
    \begin{tabular}{c | c c c}
        {Direction} &
        $\bar{\theta}_{\text{HS--O--LS}}$ ($^\circ$) &
        $\bar{\theta}_{\text{LS--O--LS}}$ ($^\circ$) &
        $\theta_{\text{LS--O--LS}}^{Bulk}$ ($^\circ$) \\
    \hline
        In-plane (\textit{xy}) & 165.21  & 163.59 & 163.23 \\
        Out-of-plane (\textit{z}) & 159.94 & 155.82 & 163.23 \\
    \end{tabular}

    \caption{Average structural parameters of CoO$_6$ octahedra in the ground state configuration. HS Co ions exhibit larger octahedral volumes and longer in-plane Co--O bond lengths compared to LS$^{(1)}$ ions (in the LS planes) and LS$^{(2)}$ ions (within the columns), while their out-of-plane bond lengths remain nearly unchanged. The average in-plane Co--O--Co bond angles are comparable to those in bulk, whereas the out-of-plane angles are noticeably reduced. The HS--O--LS$^{(1)}$ and HS--O--LS$^{(2)}$ bond angles are nearly identical, so they are not differentiated in this table.
}
    \label{tab:bonds}
\end{table}

\begin{figure}[!h]
    \centering
    \includegraphics[width=1\textwidth]{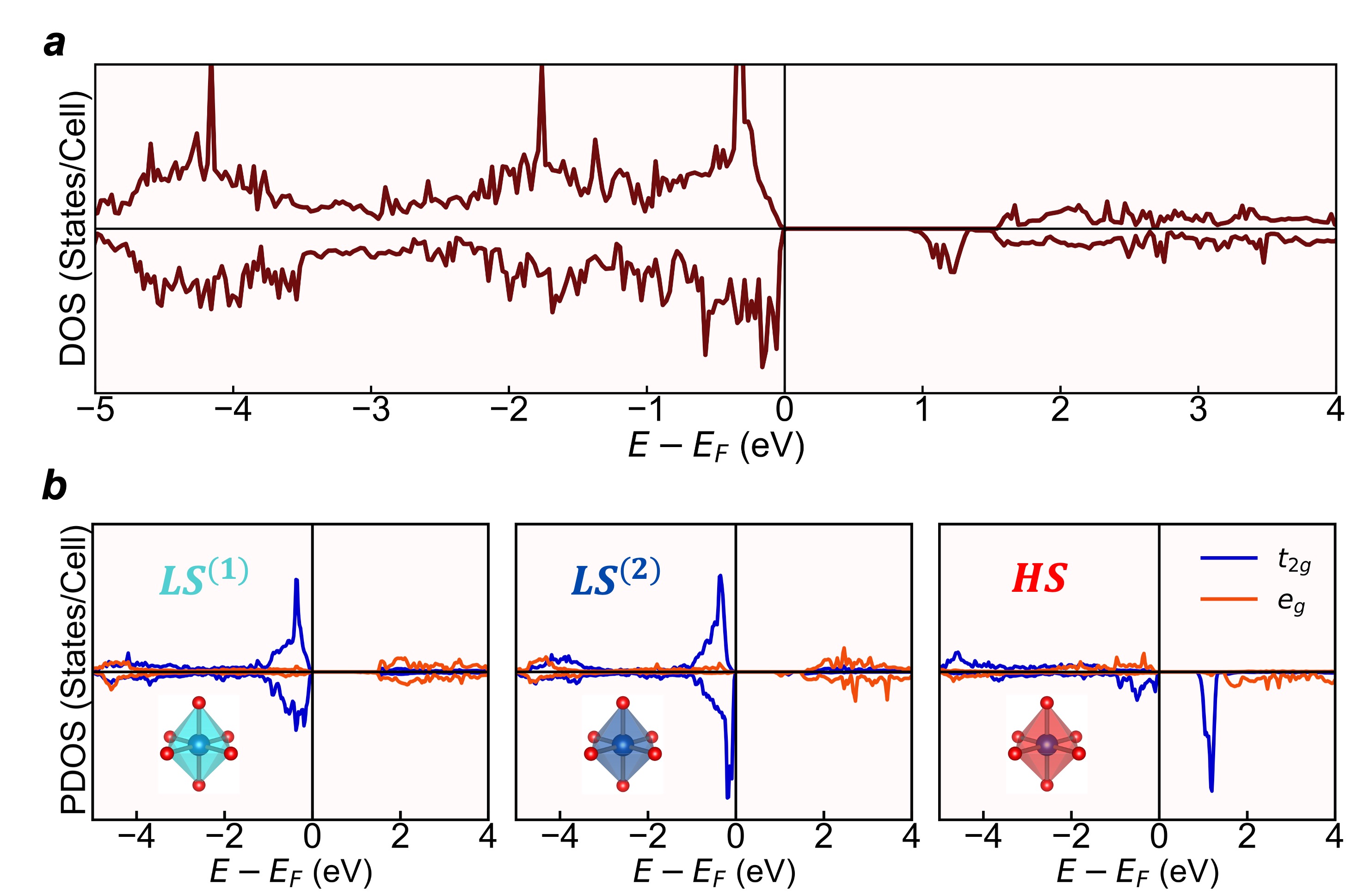}
    \caption{Density of states (DOS) and projected density of states (PDOS) for the ferromagnetic columnar model. Integration of the PDOS quantitatively confirms the spin state assignments of the Co ions. (a) Total DOS confirms the insulating nature of the system, with a band gap of 0.98~eV. (b) PDOS for LS$^{(1)}$ ions, located in the separating LS planes, shows fully occupied $t_{2g}$ orbitals. PDOS for LS$^{(2)}$ ions, positioned within the rocksalt-type HS/LS columns, also exhibits fully occupied $t_{2g}$ orbitals. PDOS for HS Co ions reveals partially occupied $e_g$ orbitals, consistent with the HS state. The results further indicate a slight charge/spin disproportionation between LS$^{(2)}$ and HS Co ions within the column.
    }
    \label{fig:PDOS}
\end{figure}

To assess the electronic structure, the density of states (DOS) was calculated for the ferromagnetic columnar model, as shown in \fig{PDOS}(a). The total DOS exhibits a band gap of approximately 0.98 eV, confirming the insulating nature of this model. The projected DOS (\fig{PDOS}(b)) further shows distinct orbital occupancies for LS$^{(1)}$, LS$^{(2)}$, and HS Co ions, illustrating the characteristic splitting of $t_{2g}$ and $e_g$ orbitals. Integration of the PDOS confirms that all LS Co ions have fully occupied $t_{2g}$ orbitals and empty $e_g$ orbitals, while the HS Co ions exhibit partial $e_g$ occupancy, providing quantitative verification of the spin state differentiation. Overall, the ferromagnetic columnar model is confirmed to be an insulating ferromagnet, with net magnetization and electronic properties consistent with experimental observations. These results establish the validity of this model as a representation of tensile-strained LaCoO$_3$ films and provide a foundation for analyzing the magnetic interactions that stabilize the ferromagnetic ordering.

\begin{figure}
     \centering
         \includegraphics[width=\textwidth]{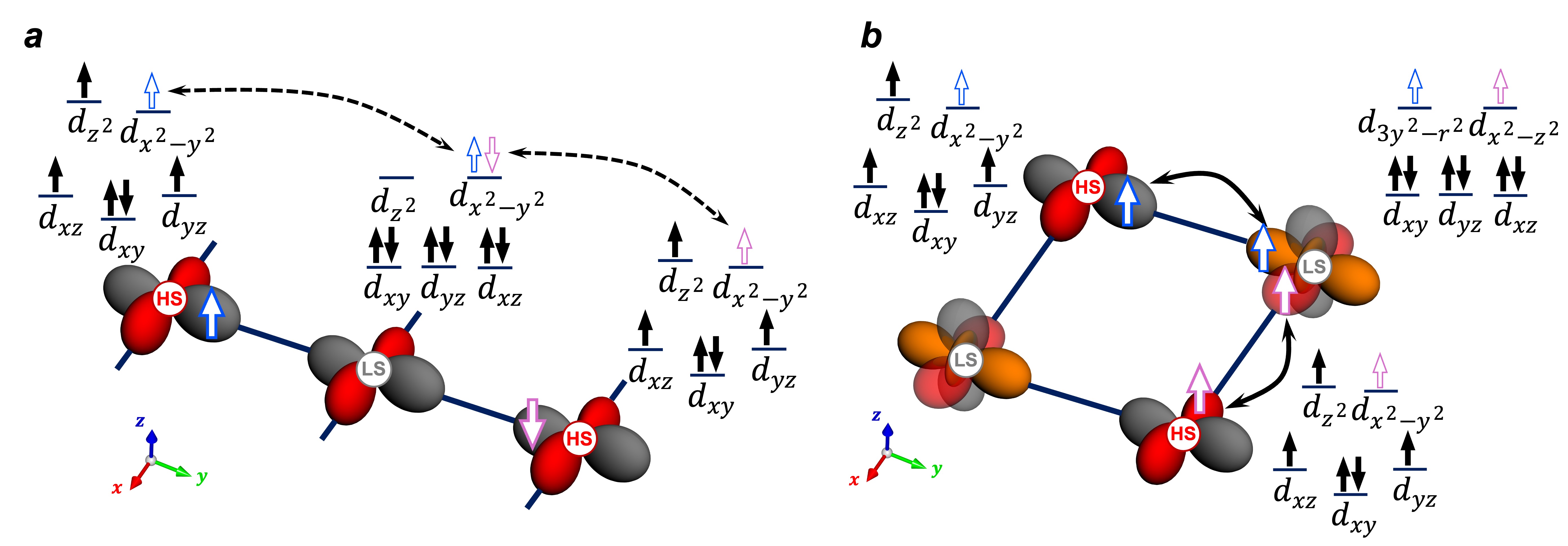}
        \caption{ Magnetic interactions within a column. (a) Antiferromagnetic superexchange pathway along a 180° bond involving a single LS Co ion situated between two HS Co ions. Only the $d_{x^2 - y^2}$ orbitals of the three Co atoms are shown; the oxygen $p_y$ orbitals, which mediate the superexchange, are omitted for clarity. Each arrow between Co atoms denotes an effective hopping process that scales as $t_{pd}^2$, where $t_{pd}$ is the Co–O hybridization parameter. (b) Ferromagnetic superexchange pathway along a 90° bond, where two HS Co ions interact via a corner LS Co ion. The pathway involves hybridization through orthogonal $e_g$ orbitals ($d_{x^2-z^2}$ and $d_{3y^2-r^2}$) on the LS Co, coupled respectively to $p_x$ and $p_y$ orbitals on the bridging oxygen atoms (not shown). Hund’s coupling on the LS Co ion favors parallel spin alignment, stabilizing a ferromagnetic interaction between the HS Co ions.}
        \label{fig:short}
\end{figure}

\begin{figure}
     \centering
         \includegraphics[width=\textwidth]{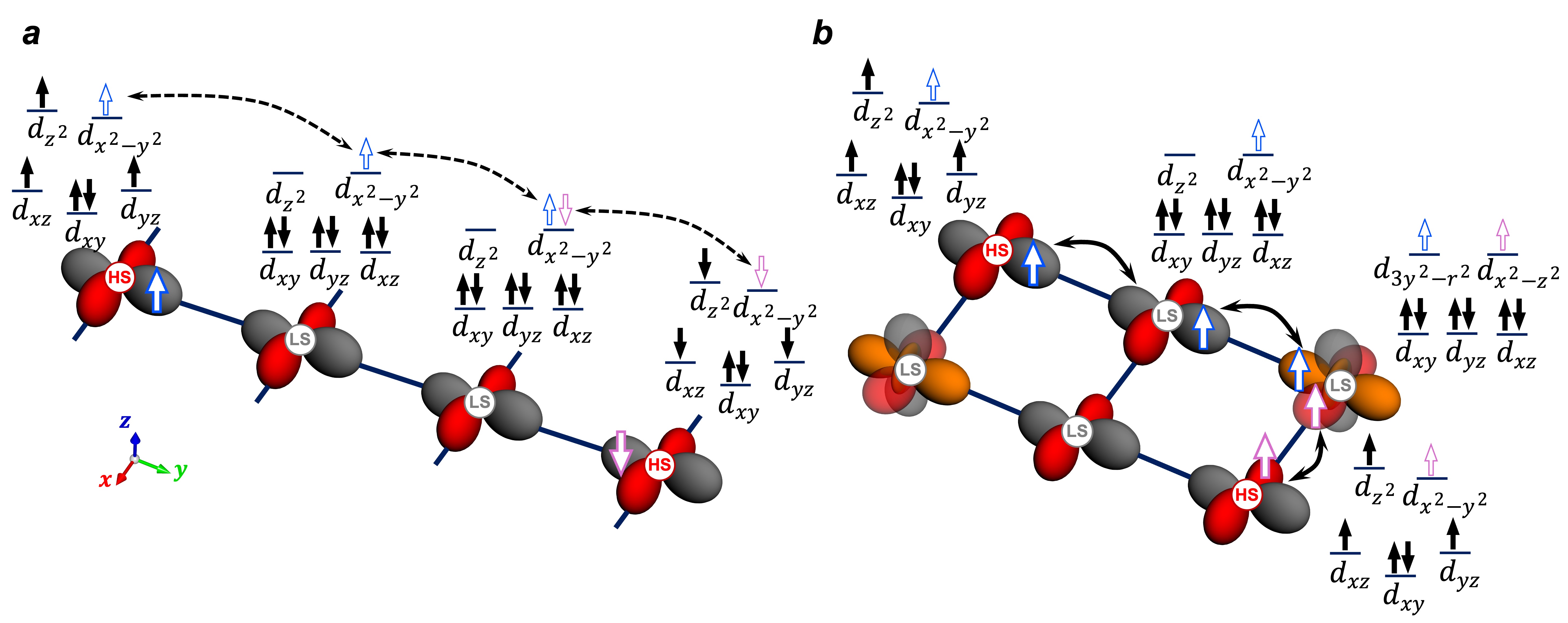}
        \caption{Magnetic interactions between columns. (a) Antiferromagnetic superexchange pathway along a longer 180° bond, involving two intermediate LS Co ions between a pair of HS Co ions. The interaction is mediated by three consecutive Co–O–Co links and arises from a virtual hopping process of $e_g$ electrons, resulting in an antiferromagnetic interaction that scales as $t_{pd}^{12}$. (b) Ferromagnetic superexchange pathway along a 90° “L-shaped” bond connecting HS Co ions in neighboring columns via two LS Co ions. Three distinct paths contribute to this interaction, increasing its overall strength.}
        \label{fig:long}
\end{figure}

The \textit{ferromagnetic columnar model} contains four distinct superexchange interactions, two within each column and two between neighboring columns, as shown in Figs.~\ref{fig:short}-\ref{fig:long}. These are: (i) a 180$^\circ$ Co(HS)--O--Co(LS)--O--Co(HS) pathway with exchange coupling denoted by $J_0$, and (ii) a 90$^\circ$ Co(HS)--O--Co(LS)--O--Co(HS) pathway with exchange coupling denoted by $J_1$; both $J_0$ and $J_1$ are within a FM column and are therefore mostly perpendicular to the substrate. In addition, (iii) a longer 180$^\circ$ Co(HS)--O--Co(LS)--O--Co(LS)--O--Co(HS) pathway with exchange coupling denoted by $J_2$, and (iv) a longer 90$^\circ$ L-shaped Co(HS)--O--Co(LS)--O--Co(LS)--O--Co(HS) pathway with exchange coupling denoted by $J_3$; both $J_2$ and $J_3$ are exchange couplings between FM columns and are therefore mostly in the plane parallel to the substrate. To quantify these magnetic couplings, the total energies of different spin configurations were mapped onto an Ising Hamiltonian of the form
\begin{equation} \label{eq:total_energy1}
E = -\left(
J_0 \sum_{\langle i,j \rangle_0} s_i s_j +
J_1 \sum_{\langle i,j \rangle_1} s_i s_j +
J_2 \sum_{\langle i,j \rangle_2} s_i s_j +
J_3 \sum_{\langle i,j \rangle_3} s_i s_j
\right),
\end{equation}
where $s_i$ represents the Ising spin variable associated with each Co site.

Assuming that all HS Co$^{3+}$ ions are equivalent, Eq.~\eqref{eq:total_energy1} can be simplified to
\begin{equation} \label{eq:total_energy2}
\Delta E = - S^2 \sum_{i=0}^{3} n_i J_i ,
\end{equation}
where $S$ is the spin of the HS Co$^{3+}$ ion ($S = 2$), $n_i$ denotes the number of corresponding superexchange interactions of type $i$, and $\Delta E$ is the total energy difference between the ferromagnetic ground state configuration and a magnetic configuration in which selected spins are flipped (see Supporting Information). By solving the resulting set of linear equations obtained from multiple spin configurations, the values of the exchange parameters $J_0$, $J_1$, $J_2$, and $J_3$ were extracted. The resulting coupling constants are summarized in \tab{secop}.

\begin{table}[ht]
    \centering
    
    \begin{tabular}{ccc}
    \hline 
        Parameter &  
        Value ($meV$) &
        Angle
        \\
    \hline
        $J_0$ &
        -5.754 $\pm$  0.327 & 180$^\circ$ \\
        
        $J_1$ &
        5.514 $\pm$  0.106 & 90$^\circ$ \\
        
        $J_2$ &
        -3.093 $\pm$  0.637 & 180$^\circ$ \\
        
        $J_3$ &
        2.736 $\pm$  0.159 & 90$^\circ$ \\

    \hline
    \end{tabular}
    \caption{Calculated superexchange coupling parameters for the \textit{ferromagnetic columnar model}. $J_0$ denotes the antiferromagnetic superexchange interaction along the out-of-plane ($z$) direction, while $J_1$ represents the ferromagnetic 90$^\circ$ superexchange interaction; both couplings act within a column. $J_2$ corresponds to the longer-range antiferromagnetic 180$^\circ$ superexchange interaction, and $J_3$ to a ferromagnetic L-shaped superexchange interaction, both operating between neighboring columns. Consistent with the discussion above, the intra-column interactions ($J_0$ and $J_1$) are stronger than the inter-column interactions ($J_2$ and $J_3$).}

    \label{tab:secop}
\end{table}

To gain insight into the mechanism underlying ferromagnetic ordering in the ground state, we analyze the superexchange interactions in spatial arrangement of HS and LS Co ions revealed by DFT; 
see Figs.~\ref{fig:short}-\ref{fig:long}.
Our goal is to understand the \textit{signs} (ferromagnetic or antiferromagnetic) and the \textit{relative strengths} of the magnetic interactions arising from different superexchange pathways between HS ions. 

We begin with the simplest case of 
a 180$^\circ$ Co(HS)--O--Co(LS)–O–Co(HS) superexchange pathway shown in Fig.~\ref{fig:short}(a).
The HS Co ion is in a $t_{2g}^4e_g^2$ configuration with $S=2$, while the LS Co ion is in a $t_{2g}^6e_g^0$ configuration with $S=0$. Superexchange arises 
from an $e_g$ electron hopping through oxygen $p_\sigma$ orbitals, specifically the $d_{x^2 - y^2}$ orbitals along the $x$-axis, which strongly overlap with the $p_y$ orbitals of the bridging O atoms. (The relevant oxygen orbitals are omitted from the figures for clarity). Hopping of $t_{2g}$ electrons is generally much weaker due to the smaller hybridization with O $p_\pi$ orbitals. In this case, it is completely suppressed by Pauli exclusion as the $t_{2g}$ orbitals of LS Co are fully occupied. The virtual hopping process involves a transient double occupancy of the $d_{x^2 - y^2}$ orbital on the LS Co ion, which is possible only when the spins on the two HS Co ions are antiparallel. This results in an antiferromagnetic interaction of the form $J_0 \mathbf{S}_1 \cdot \mathbf{S}_2$, where $\mathbf{S}_1$ and $\mathbf{S}_2$ denote the spin-2 moments of the HS Co ions.

This is essentially the same as conventional antiferromagnetic superexchange in a 180$^\circ$ bond between two transition metal (TM) ions separated by O, which scales as $t_{pd}^4$, where $t_{pd}$ denotes the TM–O hybridization matrix element. In the Co(HS)–O–Co(LS)–O–Co(HS) geometry considered here, however, the exchange interaction $J_0$ is smaller, scaling like $t_{pd}^8$. The associated energy denominators in perturbation theory depend on several material-specific parameters, including the charge transfer energy, the intra-orbital and inter-orbital Coulomb repulsions and the Hund’s coupling on the LS Co, and the crystal field splitting differences between HS and LS Co ions. A detailed discussion of how these parameters enter in the calculation can be found in \ref{app:superex}.

We note that extended 180$^\circ$ superexchange pathways involving Co(HS)-O-Co(LS)-O-Co(HS) are not present in the 
$xy$-plane of the ``columnar structure" found in the DFT, but they are present 
along the $z$-axis. The same AFM interaction $J_0$ can be derived along the $z$-axis, following the discussion above, by considering  the virtual hopping of electrons in $d_{3z^2 - r^2}$ via 
the intervening O $p_z$ orbitals.

We next turn to the 90$^\circ$ Co(HS)--O--Co(LS)--O--Co(HS) superexchange pathway shown in 
Fig.~\ref{fig:short}(b). Here, we focus on the $d_{x^2 - y^2}$ orbitals of the two HS Co ions
that couple to the $e_g$ orbitals of the intervening LS Co ion through an O $p_x$ orbital along one leg 
and an O $p_y$ orbital along the other. This geometry gives rise to two distinct 
virtual hopping processes, each critically dependent on the specific LS Co orbitals involved. 
One process favors AFM coupling, while the other promotes FM interactions. 
As we show next, the net effect of these competing contributions ultimately favors a FM superexchange in this 90$^\circ$ configuration.

(a) The first process involves the delocalization of antiparallel spins on the HS ions with an intermediate virtual state with a doubly occupied $d_{x^2 - y^2}$ orbital on the LS Co. The details of the energetics, specifically, the energy denominators involved in the perturbation theory, are relegated to \ref{app:superex}. This analysis closely parallels the 180$^\circ$ superexchange discussed above and results in an AFM interaction.

(b) The second process corresponds to the delocalization of parallel-spin electrons on the two HS Co ions.
By Pauli exclusion, the intermediate state in the process necessarily involves two {\em orthogonal} $e_g$ orbitals on the corner LS Co ion. We choose these two orbitals to be specific 
linear combinations, $d_{x^2 - z^2}$ and $d_{3y^2 - r^2}$, that can hybridize 
with the O $p_x$ and $p_y$ orbitals on the two legs of the 90$^\circ$ bond geometry;
see Fig.~\ref{fig:short}(b). The key point here is that the energy of the intermediate state on the LS Co is lower by Hund’s coupling $J_H$ relative to case (a). The reduced energy denominator in case (b) relative to (a) leads to a net FM superexchange. 

A general expression for the the net FM superexchange interaction $J_1 \mathbf{S}_1 \cdot \mathbf{S}_2$ 
with $J_1 < 0$ along a 90$^\circ$ pathway is derived in  \ref{app:superex}. 
For $\alpha={J_H}/{U_{\rm eff}}\lesssim 1$, this simplifies to
\begin{equation}
    J_{1}= -\,\,\frac{2\alpha}{3}\,J_0.
\end{equation}
Here $J_0$ is the 180$^\circ$ AFM interaction derived above, and $\alpha$ is the
ratio of the Hund's coupling $J_H$ on LS Co to $U_{\rm eff}= U' + 2\Delta/3$,
where $U'$ is the inter-orbital Coulomb repulsions on LS Co and 
$\Delta$ the crystal field splitting difference between HS and LS Co ions.
We have included a factor of 2 enhancement in the expression for $J_1$ because, for the
HS CO ions located at opposite corners of a square, there are 
two different $90^\circ$ pathways through the two LS ions located on the other 
diagonal of the square. 

It is well known from the classic work of Goodenough, Kanamori, and Anderson
that $90^\circ$ bonds, between two transition metal ions with oxygen at the corner, lead to 
FM superexchange~\cite{khomskii_transition_2014}. In that case, the Hund's coupling between electrons in two orthogonal 
$p_\sigma$ oxygen orbitals stabilizes FM exchange. Our analysis is very similar, with the 
following differences. 
(i) the Co-O-Co bonds are all $180^\circ$, but we have $90^\circ$ Co(HS)--O--Co(LS)--O--Co(HS) bonds. 
(ii) This forces us to look at a linear combination of the standard $e_g$ orbitals on the LS Co atom, 
which can hybridize with the relevant $p_\sigma$ O orbitals and thus with the HS Co's. These are 
the $d_{x^2 - z^2}$ and $d_{3y^2 - r^2}$ orbitals on LS Co in the analysis above.
(iii) It is the Hund's coupling between electrons in these two orthogonal $e_g$ orbitals on LS Co that stabilizes FM exchange.
 
 The AFM coupling $J_0$ along the $z$-axis and the ferromagnetic exchange $J_1$ in the $xy$, $xz$ and $yz$ planes 
 describe the magnetic interactions within a column. We next need to look at the longer range couplings 
 between the columns, before turning to the question of the magnetic ground state of the entire system.
 
To understand the interactions between columns, we need to look at superexchange pathways of the form 
Co(HS)–O–Co(LS)–O–Co(LS)–O–Co(HS). The two inter-column interactions that arise are (1) a 180$^\circ$ linear configuration, shown in Fig.~\ref{fig:long}(a), and (2) an L-shaped 90$^\circ$ configuration, shown in 
Fig.~\ref{fig:long}(b).

The 180$^\circ$ interaction (Fig.~\ref{fig:long}(a)) results in an AFM interaction $J_{2} > 0$
between columns, which scales as $t_{pd}^{12}$ weaker than the $t_{pd}^{8}$ interactions within a column. The AFM nature of $J_2$ arises as before: Pauli exclusion implies electrons with parallel spins cannot occupy the same orbital in intermediate virtual states.

The L-shaped configuration (Fig.~\ref{fig:long}(b)) leads to a FM interaction $J_{3} < 0$. This is
again stabilized by selecting two orthogonal $e_g$ orbitals on the corner LS Co ion, so that the Hund’s coupling energy gained by parallel spins reduces the energy denominator. Now, we must take into account a 
factor of three enhancement in the FM interaction because of three distinct exchange paths connecting the two HS Co ions in this geometry (the two L-shaped paths and a ``zig-zag" path). 
Nevertheless, the FM $|J_3|$ will be smaller than AFM $|J_2|$ by a factor of $\alpha = J_H/U_{\rm eff}$,
as in the case of intra-column interactions.

We have checked using conjugate gradient minimization of a classical Heisenberg Hamiltonian with the 
HS Co ions arranged as given by DFT, and retaining all the interactions $J_0$ through $J_3$, that the 
the ground state has ferromagnetic long range order. The combination of the FM interactions
$J_1$ within the columns and $J_3$ between columns, together with the multiplicity of FM neighbors, 
is able to stabilize FM long range order despite the presence of competing AFM interactions $J_0$ and $J_2$.

We conclude with a comment on analyzing this system at finite temperature and computing its transition temperature $T_c$. The difficulty here stems from the fact that the spin state of the Co ions is itself
a function of temperature. Even in bulk LaCoO$_3$, which has a diamagnetic ground state with all Co ions in the LS configuration ($S=0$), the spin-state population changes with temperature \cite{goodenough_interpretation_1958}: as $T$ increases, the concentration of HS Co$^{3+}$ rises rapidly at the expense of LS Co$^{3+}$, reaching a 50:50 mixture near 110 K. This evolution reflects the higher entropy of the $S\neq 0$ states relative to the unique $S=0$ state. As a result, such spin-state changes complicate any finite-temperature analysis based solely on the $T=0$ spatial ordering of HS and LS Co ions.

\newpage

\section{Conclusions}

In summary, a microscopic mechanism for the ferromagnetic insulating state in tensile-strained, stoichiometric LaCoO$_3$ thin films has been established using first-principles calculations. By explicitly isolating the role of epitaxial strain, the analysis shows that sufficiently large tensile strain can drive a spin state transition of Co$^{3+}$ ions from the low-spin to the high-spin state, even in the absence of oxygen vacancies. When strain corresponding to a SrTiO$_3$ substrate is imposed, the energetically favored ground state consists of ferromagnetically aligned columnar regions with rocksalt-type high-spin/low-spin ordering, separated by planes of low-spin Co ions. Electronic structure calculations further confirm the insulating nature of this configuration. The calculated superexchange coupling parameters indicate that 180$^\circ$ Co-O-Co pathways through low-spin Co$^{3+}$ ions are antiferromagnetic, whereas 90$^\circ$ pathways favor ferromagnetic interactions. Accounting for all such interactions together, ferromagnetic couplings are dominant, providing the energetic basis for the stabilization of the columnar magnetic ground state.

Overall, this work establishes epitaxial strain as a sufficient driving force for ferromagnetic insulating behavior in LaCoO$_3$ thin films and offers a consistent microscopic picture that reconciles electronic structure, lattice distortion, and magnetic interactions. These insights provide a foundation for strain-engineered control of spin states and magnetism in correlated oxide heterostructures.

\section*{Acknowledgments} This work was  supported by the Center for Emergent Materials (CEM), a National Science Foundation MRSEC under NSF Award Number DMR-2011876. Computational resources were provided through the Ohio Supercomputer center.

\section*{Data Availability} The data supporting this study’s findings are available within the article.

\appendix

\section{Derivation of Superexchange Interactions}\label{app:superex}

In this Appendix, we estimate the magnitudes and signs of the exchange couplings generated by the superexchange pathways 
discussed in the main text. We will consider the HS-LS-HS geometries shown in 
Figs.~\ref{fig:short}(a) and \ref{fig:short}(b) where
the two magnetic ions are HS Co ($t_{2g}^4 e_g^2$, spin $S=2$) and the intermediate ion is LS Co ($t_{2g}^6 e_g^0$, spin $S=0$). 
The $e_g$ electrons on HS Co hop into the empty $e_g$ orbitals of LS Co via oxygen $p_\sigma$ orbitals
(which are omitted in the Figures for clarity). We can ignore the hopping of the $t_{2g}$ electrons; these are generally weaker 
due to the smaller hopping amplitudes via oxygen $\pi$ orbitals, but in fact these completely suppressed due to 
Pauli blocking on the LS site.

We write the Hamiltonian as $H=H_0+V$, where $H_0$ contains the on-site interaction energies, while the perturbation
$V$ describes the Co--O hopping with amplitude $t_{pd}$. We consider processes where the electron virtually moves from the HS Co to the LS Co $e_g$ manifold through oxygen, and then returns. To second order in V, we can eliminate O to obtain an effective HS--LS hopping amplitude
\begin{equation}
t \;\sim\; \frac{t_{pd}^2}{\Delta_{CT}},
\label{eq:teff_leg}
\end{equation}
where $\Delta_{CT}$ is the charge-transfer energy.
The HS--HS exchange interaction then arises in \emph{fourth order} in this effective hopping $t$, or 
equivalently \emph{eighth order} in $t_{pd}$.

Let us first look at the $180^\circ$ pathway shown in Fig.~\ref{fig:short}(a).
To keep track of the virtual processes, it is convenient to label the relevant charge configurations by the occupancies on the three Co sites (HS--LS--HS) as shown in Table~\ref{tab:states_180}. The ground state is denoted by the label $S1$ in this Table, while 
intermediate states that arise in the perturbation theory involve either one (states $S2$ and $S3$) or two electrons (in state $S4$) that have been transferred to the LS $e_g$ orbitals. In the $180^\circ$ geometry, the electrons originating from the two HS ions must hop onto the \emph{same} LS orbital. Thus, the doubly occupied LS state $S4$ can arise only from an antiparallel spin configuration on the HS sites. This, of course, is the microscopic origin of the antiferromagnetic exchange. 

The energies of the intermediate states, measured relative to the ground state, for the $180^\circ$ geometry are shown in Table~\ref{tab:states_180}. Here $U$ is the intra-orbital Hubbard interaction, $U'$ is the inter-orbital Coulomb energy, 
$J_H$ is the (ferromagnetic) Hund's coupling and $\Delta$ is the difference between LS and HS crystal-field splittings.

\begin{table}[t]
\caption{Charge configurations distinguished by the occupancies on the HS--LS--HS ions and their energies for the $180^\circ$ geometry in the AFM alignment. Here $\Delta$ is the difference between LS and HS crystal-field splittings, $U$ and $U'$ are the intra-orbital and inter-orbital Coulomb interactions, respectively, and $J_H$ is Hund's coupling.}
\label{tab:states_180}

\vspace{0.3em}
\hrule
\vspace{0.15em}
\hrule
\vspace{0.5em}

\centering
\begin{tabular}{c c c c c}
State & HS Co$^{3+}$ & LS Co$^{3+}$ & HS Co$^{3+}$ & Energy \\
\hline
$S1$ & $t_{2g}^4e_g^2$ & $t_{2g}^6e_g^0$ & $t_{2g}^4e_g^2$ & $0$ \\
$S2$ & $t_{2g}^4e_g^1$ & $t_{2g}^6e_g^1$ & $t_{2g}^4e_g^2$ & $U'+J_H+\Delta$ \\
$S3$ & $t_{2g}^4e_g^2$ & $t_{2g}^6e_g^1$ & $t_{2g}^4e_g^1$ & $U'+J_H+\Delta$ \\
$S4$ & $t_{2g}^4e_g^1$ & $t_{2g}^6e_g^2$ & $t_{2g}^4e_g^1$ &
$2U'+U+2J_H+2\Delta \quad (\uparrow\downarrow)$ \\
\end{tabular}

\vspace{0.5em}
\hrule
\vspace{0.15em}
\hrule
\end{table}

To calculate the exchange coupling, we write the fourth-order correction to the energy of an (unperturbed) ground state $\ket{n^{(0)}}$ in the standard Rayleigh--Schr\"odinger perturbation theory. This is given by
\begin{align}
E_n^{(4)} &=
\sum_{k_2,k_3,k_4}
\frac{V_{n k_4}V_{k_4 k_3}V_{k_3 k_2}V_{k_2 n}}
{E_{n k_2}\,E_{n k_3}\,E_{n k_4}}
-
\sum_{k_2,k_4}
\frac{\left|V_{n k_4}\right|^2\left|V_{n k_2}\right|^2}
{E_{n k_4}^2\,E_{n k_2}}
\nonumber\\
&\hspace{1.2cm}
-2\sum_{k_3,k_4}
\frac{V_{n k_4}V_{k_4 k_3}V_{k_3 n}\,V_{nn}}
{E_{n k_3}^2\,E_{n k_4}}
+
\sum_{k_4}
\frac{\left|V_{n k_4}\right|^2\,V_{nn}^2}
{E_{n k_4}^3}.
\label{eq:E4_RS_full}
\end{align}
where we use the notation
\begin{equation}
V_{nm}\equiv \bra{n^{(0)}}V\ket{m^{(0)}},\qquad
E_{nm}\equiv E_n^{(0)}-E_m^{(0)},
\end{equation}
In our problem the perturbation $V$ describes HS--LS hopping, and $V_{nn}=0$ for the unperturbed charge configurations and thus Eq.~\eqref{eq:E4_RS_full} simplifies to
\begin{equation}
E_n^{(4)} =
\sum_{k_2,k_3,k_4}
\frac{V_{n k_4}V_{k_4 k_3}V_{k_3 k_2}V_{k_2 n}}
{E_{n k_2}\,E_{n k_3}\,E_{n k_4}}
-
\sum_{k_2,k_4}
\frac{\left|V_{n k_4}\right|^2\left|V_{n k_2}\right|^2}
{E_{n k_4}^2\,E_{n k_2}}.
\label{eq:E4_RS_simplified}
\end{equation}

For HS--LS--HS superexchange, each $V_{nm}\sim t$ whenever $\ket{m^{(0)}}$ and $\ket{n^{(0)}}$ differ by a single allowed hop. 
The important spin-dependent fourth-order processes are ``closed loops" that start in the ground state configuration $S1$ of Table~\ref{tab:states_180}, visit states $S2$ or $S3$, then pass through $S4$, back to $S2$ or $S3$, before returning to $S1$. 
There are four equivalent ways to realize such a loop -- two choices for which HS electron hops first, and two choices for the return sequence -- giving an overall multiplicity factor of $4$.

The second term in Eq.~\eqref{eq:E4_RS_simplified} is a product of two independent second-order processes and depends only 
on the $S2$ and $S3$ denominators, without ever involving $S4$. It is thus identical for the FM and AFM configurations 
of the two HS Co ions. Since the exchange coupling is obtained from the energy difference $E_{\rm FM}-E_{\rm AFM}$, 
this second term does not contribute to the exchange.

In contrast, the first term contains the ``closed loop" with the intermediate state $S4$ of the two-electron LS and is therefore sensitive to spin alignment. In the $180^\circ$ pathway, the FM configuration cannot access $S4$ because double occupancy of the same LS orbital by two parallel-spin electrons is forbidden by the Pauli exclusion principle. Hence, the first term vanishes for the FM configuration, and the antiferromagnetic exchange $J_0$ is determined just by the first term evaluated in the AFM configuration. Using the intermediate state energies shown in Table~\ref{tab:states_180}, we obtain
\begin{equation}\label{eq:J0}
\begin{aligned}
    &\qquad \qquad \qquad H =J_0 \, \, \mathbf{S}_1 \cdot \mathbf{S}_2\\
    J_0 \;&=\; \,\frac{4t^4}{\bigl(U' + J_H + \Delta\bigr)^2\bigl(3U' + 4J_H + 2\Delta\bigr)}.
\end{aligned}
\end{equation}
where we used $U=U'+2J_H$~\cite{khomskii_transition_2014} for the intra-orbital Coulomb repulsion.

\begin{table}[t]
\caption{Charge configurations and energies for the $90^\circ$ geometry. In the two-electron intermediate state $S4$, the two transferred electrons occupy \emph{orthogonal} LS $e_g$ orbitals and thus two spin configurations are possible, as indicated. Here $\Delta$ is the difference between LS and HS crystal-field splittings, $U'$ is the inter-orbital Coulomb interaction, and $J_H$ is Hund's coupling.}
\label{tab:states_90}

\vspace{0.3em}
\hrule
\vspace{0.15em}
\hrule
\vspace{0.5em}

\centering
\begin{tabular}{c c c c c}
State & HS Co$^{3+}$ & LS Co$^{3+}$ & HS Co$^{3+}$ & Energy \\
\hline
$S1$ & $t_{2g}^4e_g^2$ & $t_{2g}^6e_g^0$ & $t_{2g}^4e_g^2$ & $0$ \\
$S2$ & $t_{2g}^4e_g^1$ & $t_{2g}^6e_g^1$ & $t_{2g}^4e_g^2$ & $U'+J_H+\Delta$ \\
$S3$ & $t_{2g}^4e_g^2$ & $t_{2g}^6e_g^1$ & $t_{2g}^4e_g^1$ & $U'+J_H+\Delta$ \\
$S4$ & $t_{2g}^4e_g^1$ & $t_{2g}^6e_g^2$ & $t_{2g}^4e_g^1$ &
$\begin{array}{c}
3U' + J_H + 2\Delta \quad (\uparrow\uparrow)\\
3U' + 2J_H + 2\Delta \quad (\uparrow\downarrow)
\end{array}$ \\
\end{tabular}

\vspace{0.5em}
\hrule
\vspace{0.15em}
\hrule
\end{table}
Next, we turn to the $90^\circ$ pathway; see Fig.~\ref{fig:short}(b).
In contrast to the $180^\circ$ case, the $S4$ intermediate state now involves two electrons occupying \emph{orthogonal} LS $e_g$ orbitals. Therefore, both spin configurations $(\uparrow\uparrow)$ and $(\uparrow\downarrow)$ are allowed, and consequently \emph{both} FM and AFM alignments contribute to the first term in Eq.~\eqref{eq:E4_RS_simplified}. 
As before, the second term in Eq.~\eqref{eq:E4_RS_simplified} depends only on the one-electron 
charge-transfer denominators and is identical for FM and AFM configurations; 
it therefore cancels when we take the difference $E_{\rm FM}-E_{\rm AFM}$ to extract the exchange coupling.

We summarize in Table~\ref{tab:states_90} the energies of the states $S1$ through $S4$ for the $90^\circ$ case.
The difference between FM and AFM alignments arises entirely from the energy of the two-electron intermediate state $S4$. For the $90^\circ$ geometry, the Hund's coupling lowers the energy of the parallel-spin configuration. 
This leads to $E_{\rm FM}^{(4)} < E_{\rm AFM}^{(4)}$, and hence the net exchange is ferromagnetic.

Using the energies listed in Table~\ref{tab:states_90}, we find that the $90^\circ$ exchange is
\begin{equation}
\begin{aligned}
    &\qquad \qquad \qquad H =J_1 \, \, \mathbf{S}_1 \cdot \mathbf{S}_2,\\[4pt]
    J_1 \;&=\; -\,2\times \frac{4t^4}{\bigl(U' + J_H + \Delta\bigr)^2}
    \frac{J_H}{\bigl(3U' + 2J_H + 2\Delta \bigr)\bigl(3U' + J_H + 2\Delta \bigr)}.
\end{aligned}
\label{eq:J1}
\end{equation}
Here the factor of 2 arises from the geometry of two HS ions located at diagonally opposite corners of a 
square, so that there are two different $90^\circ$ pathways through LS ions (on the other diagonal of the square). 

Let us finally look at the relative magnitudes of the $90^\circ$ FM and $180^\circ$ AFM exchange couplings. 
Using Eqs. \eqref{eq:J0}-\eqref{eq:J1}, one obtains
\begin{equation}
J_1 \;=\; -\,2J_0
    \frac{J_H \bigl(3U' + 4J_H + 2\Delta\bigr)}{\bigl(3U' + 2J_H + 2\Delta \bigr)\bigl(3U' + J_H + 2\Delta \bigr)}.\label{eq:J1_again}
\end{equation}
Using $U_{\rm eff}\equiv U' + 2\Delta/3$ and defining $\alpha \equiv J_H/U_{\rm eff}$, we find
\begin{equation}
    J_1=-\frac{2J_0}{3}\frac{\alpha(1+4\alpha/3)}{(1+2\alpha/3)(1+\alpha/3)}.
\end{equation}
To get a feel for this expression, we look at the regime $\alpha \ll 1$, where we obtain
\begin{equation}
J_1 \approx -\frac{2}{3}J_0\,\alpha + \cdots
\label{eq:A9}
\end{equation}
Thus, in the $\alpha \ll 1$ limit, the $90^\circ$ ferromagnetic exchange is smaller than the $180^\circ$ antiferromagnetic exchange by a factor of $\alpha = J_H/U_{\mathrm{eff}}$. We further find that this correction remains tiny even up to $\alpha \sim 1$.

\clearpage
\appendix
\section*{Supplementary Material}
\setcounter{figure}{0}
\setcounter{table}{0}
\renewcommand{\thefigure}{S\arabic{figure}}
\renewcommand{\thetable}{S\arabic{table}}
\renewcommand*{\fig}[1]{Supplementary Figure~\ref{fig:#1}}
\renewcommand*{\tab}[1]{Supplementary Table~\ref{tab:#1}}
\section{Strain Effects on a Single Co Ion}

\begin{figure}[!h]
    \centering
    \includegraphics[width=0.6\textwidth]{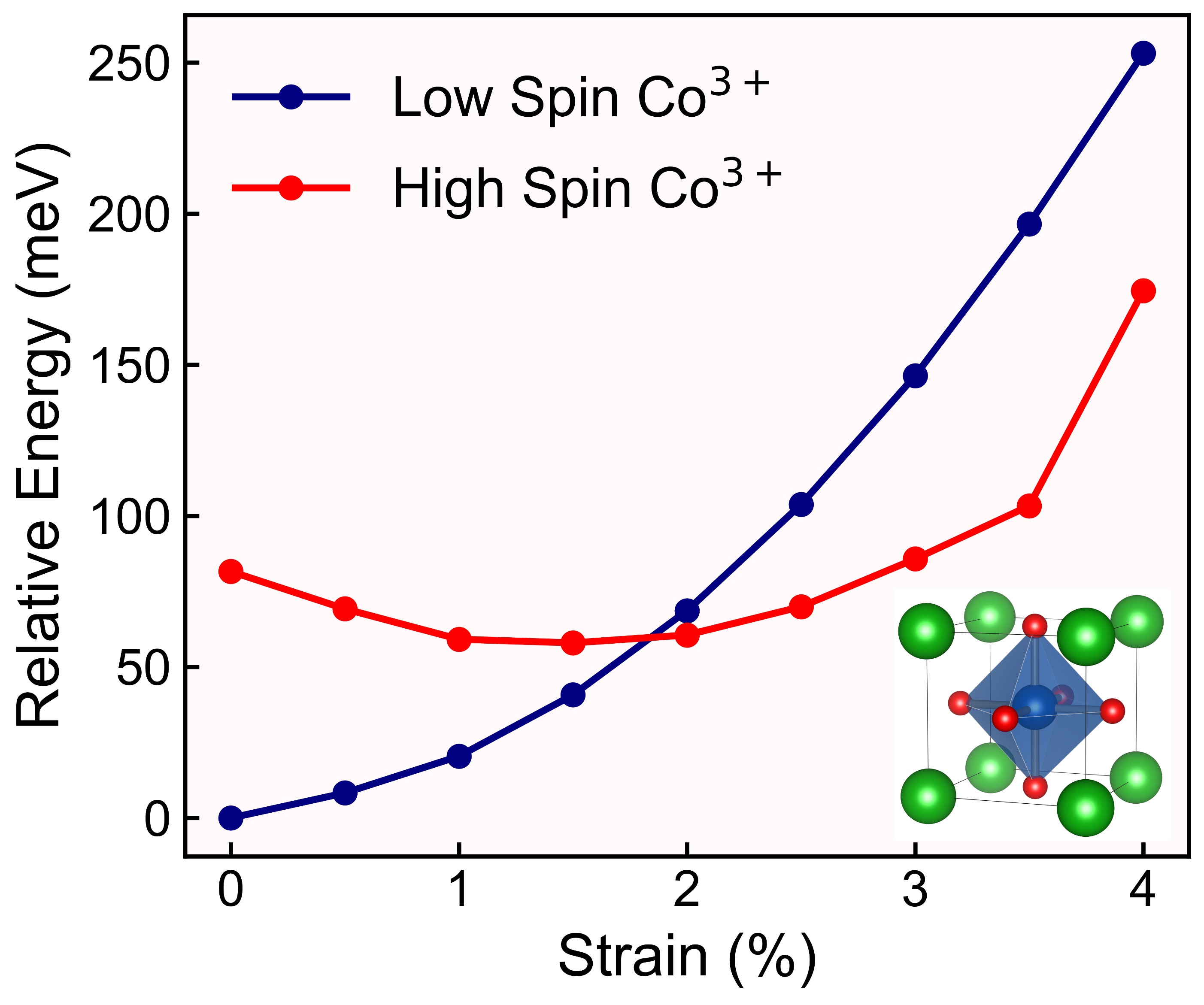}
    \caption{
    Energy difference between LS and HS states of a single Co ion in LaCoO$_3$ as a function of applied tensile strain. The HS state becomes energetically favorable above 1.8\% strain, demonstrating that epitaxial strain can induce a spin state transition. The intermediate spin (IS) state is not considered due to symmetry equivalence in the x and y directions.
    }
    \label{fig:hsls}
\end{figure}

To examine the effect of epitaxial strain on the spin state of an individual Co ion, biaxial in-plane tensile strain in the range of 0--4\% was applied to the LaCoO$_3$ unit cell. For each strain value, the total energies of the low spin (LS) and high spin (HS) configurations were calculated and compared, as shown in \fig{hsls}. The intermediate spin (IS) state was not considered, as the equivalence of the in-plane $x$ and $y$ directions under biaxial strain disfavors stabilization of an IS configuration. The results show that above approximately 1.8\% epitaxial tensile strain, the HS Co$^{3+}$ ion becomes energetically favorable relative to the LS state. This behavior is consistent with a strain-induced reduction of the crystal field splitting, which eventually becomes smaller than Hund’s exchange coupling, thereby driving a spin-state transition. These results demonstrate that sufficiently large epitaxial tensile strain alone can induce a transition from the LS to the HS state.

\section{Total Number of Configurations}

\begin{figure}[!h]
    \centering
    \includegraphics[width=1\textwidth]{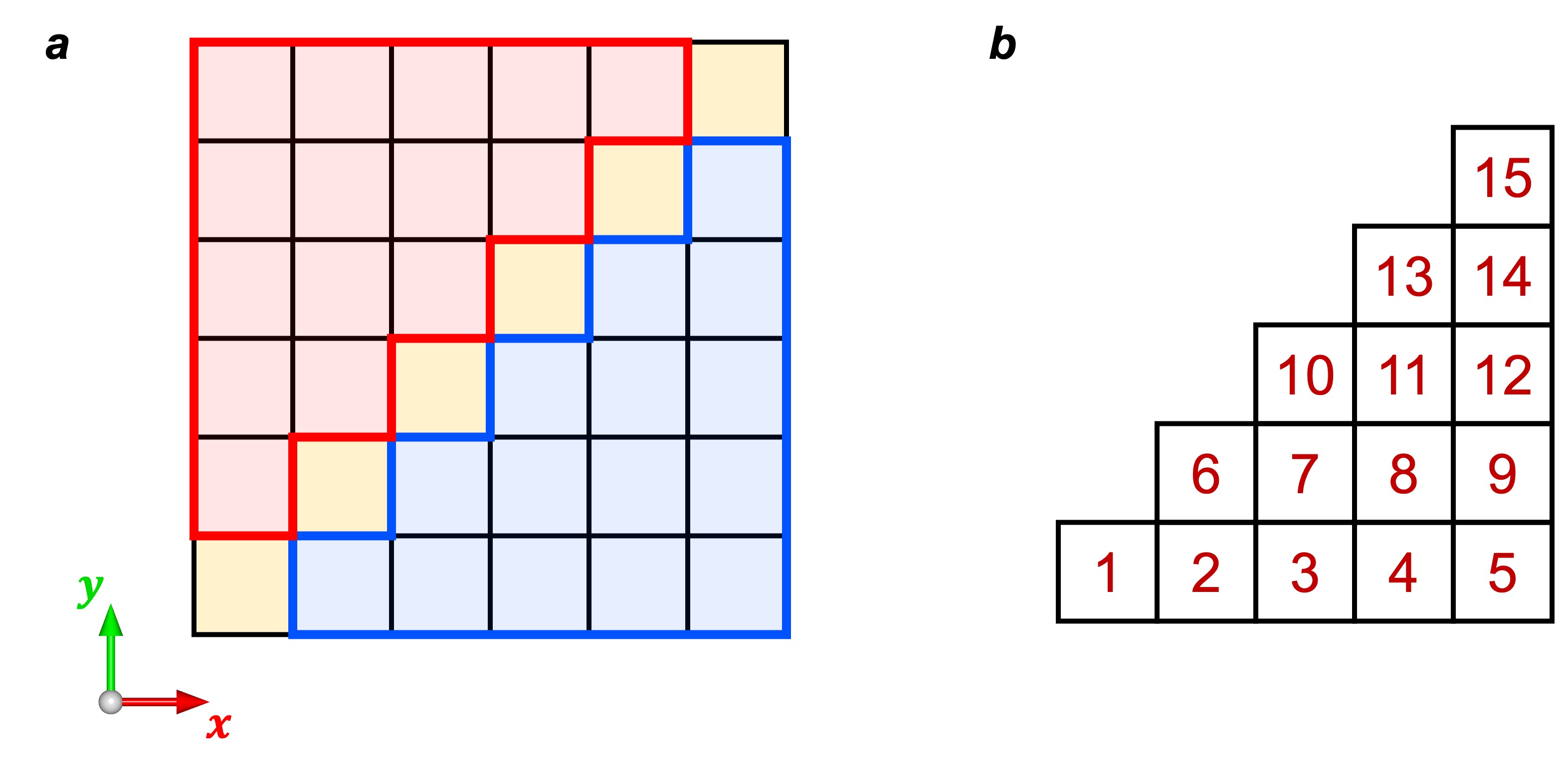}
    \caption{
    Square lattices illustrating the arrangement of Co ions. (a) A $6 \times 6$ square grid representing the Co ions in a single layer, with the irreducible unit highlighted in blue. (b) The irreducible grid is shown, with individual squares labeled from 1 to 15 for reference.
    }
    \label{fig:configs}
\end{figure}

To determine the total number of possible configurations, we begin with a $6 \times 6$ square lattice as illustrated in \fig{configs}(a). Because we wish to preserve symmetry along both the X and Y directions, it is sufficient to consider only the 12 independent squares indicated in \fig{configs}(b), which must be specified to define the full configuration.  

Each site can be in one of three spin states: high spin state up (HS$\uparrow$), high spin state down (HS$\downarrow$), or low spin state (LS), yielding a total of $3^{15} = 14,348,907$ possible configurations.  

Next, to identify the physically meaningful configurations, we impose the Goodenough--Kanamori rules. The two main rules are as follows:  

\begin{enumerate}
    \item Parallel alignment of two neighboring HS ions is forbidden in both rows and columns.
    \item Diagonal neighbors must have the same spin sign.
\end{enumerate}

To implement these rules systematically, we represent the three states numerically as $-1$ (HS$\downarrow$), $+1$ (HS$\uparrow$), and $0$ (LS). Under this representation:

\begin{itemize}
    \item Rule 1 requires that the sum of any two neighboring spins in a row or column cannot be $+2$ or $-2$.
    \item Rule 2 requires that diagonal neighbors must not sum to 0 unless at least one of them is LS (i.e., diagonal HS pairs of opposite spins are forbidden).
\end{itemize}

\begin{figure}[!h]
    \centering
    \includegraphics[width=1\textwidth]{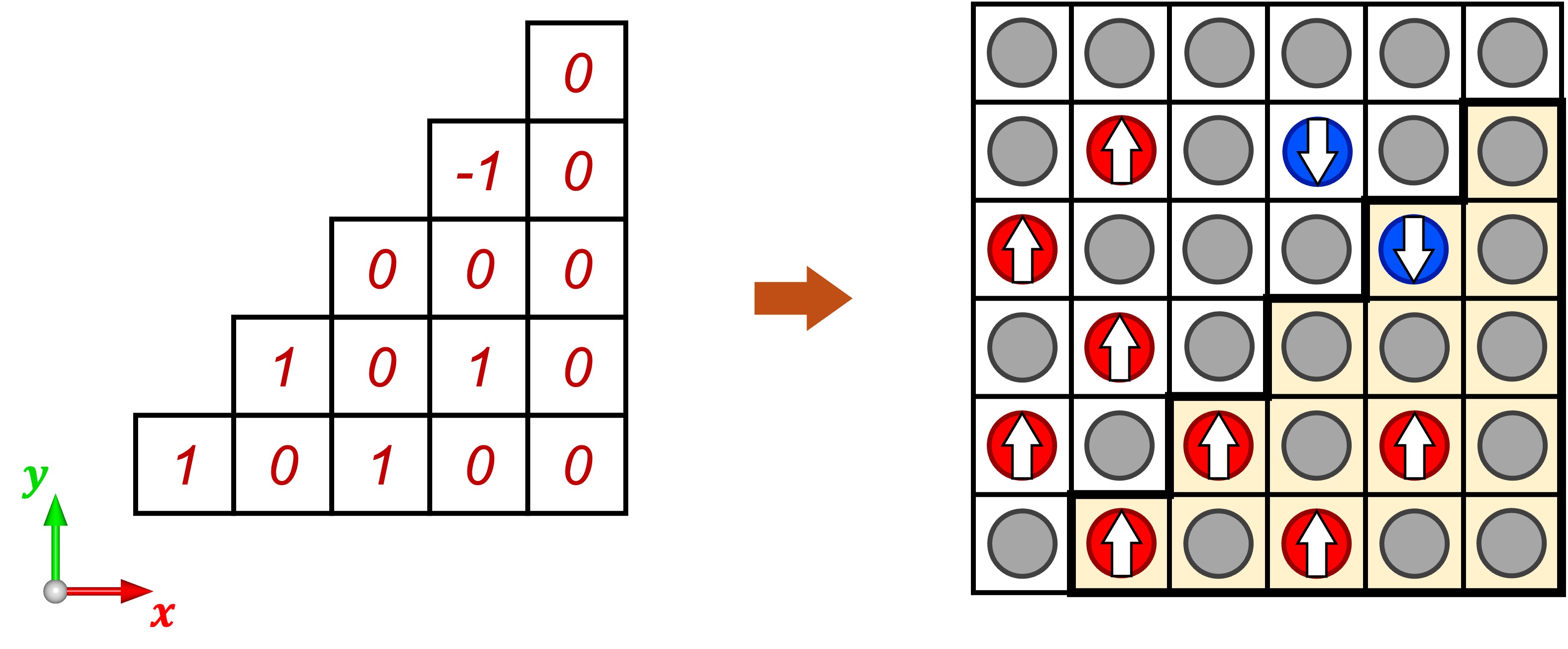}
    \caption{
    An example of a valid configuration shown both in the numerical format ($-1$, $0$, $1$) and in the corresponding spin states (HS$\uparrow$, HS$\downarrow$, and LS).
    }
    \label{fig:examp}
\end{figure}

By applying these constraints to the initial set of configurations, the number of valid configurations is reduced to $3213$ (comprising $3212$ configurations plus the single configuration with all LS ions). However, because all $15$ squares are equivalent under the periodic boundary conditions, and because both a configuration and its spin-inverted counterpart are included in this count, we divide by $30 = 2 \times 15$ to account for these symmetries. Moreover, additional random simulations indicate that placing two HS ions adjacent to each other in a row or column is energetically unfavorable. This yields a total of $95$ unique configurations. These configurations were generated by representing each combination as a $1 \times 15$ array. For example, one of the valid configurations is given by \([1, 0, 1, 0, 0, 1, 0, 1, 0, 0, 0, 0, -1, 0, 0]\), as illustrated in \fig{examp}. We also note that the magnetic arrangement in the second layer is directly related to that of the first layer in order to preserve the overall symmetry.

\section{Energy Comparison of Magnetic Configurations}

\begin{figure}[!h]
    \centering
    \includegraphics[width=0.85\textwidth]{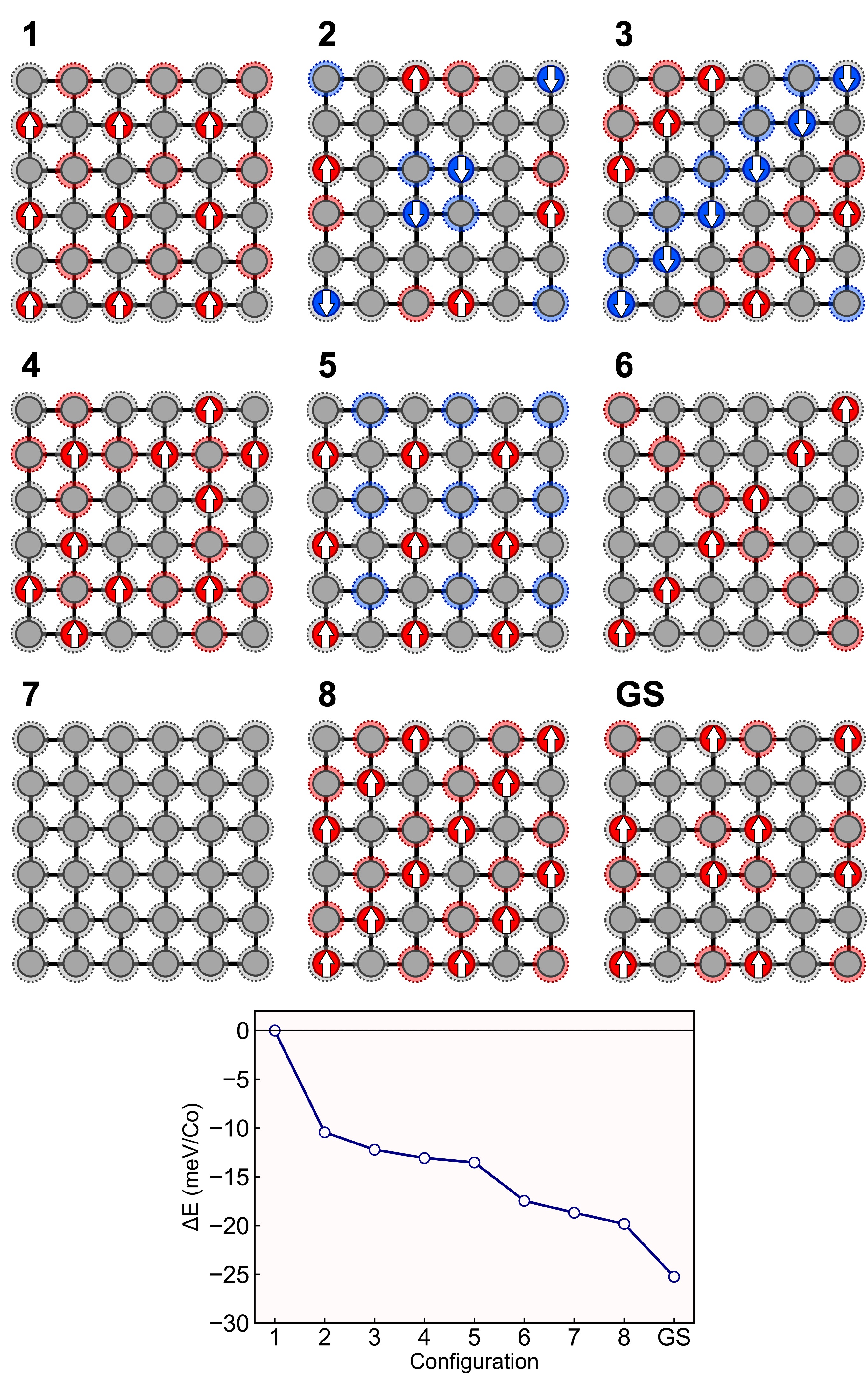}
    \caption{
    Relative total energies of eight high-symmetry magnetic configurations considered in the $3\times3\times1$ supercell. The red, blue, and gray colors represent HS ($\uparrow$), HS ($\downarrow$), and LS Co ions, respectively. The ionic arrangement in the top layer is indicated by solid lines, and the bottom layer is shown using dashed lines.
    }
    \label{fig:dE}
\end{figure}

The relative total energies of the eight high-symmetry magnetic configurations in the $3\times3\times1$ supercell were calculated and are shown in \fig{dE}. These configurations were chosen because they preserve the highest possible symmetry of the supercell and are physically meaningful. In total, 95 symmetry-inequivalent magnetic configurations were examined; however, all configurations not shown in \fig{dE} lie significantly higher in energy and do not compete with the ground state configuration. Therefore, only the energetically relevant high-symmetry configurations are discussed here.

\begin{figure}[!h]
    \centering
    \includegraphics[width=1\textwidth]{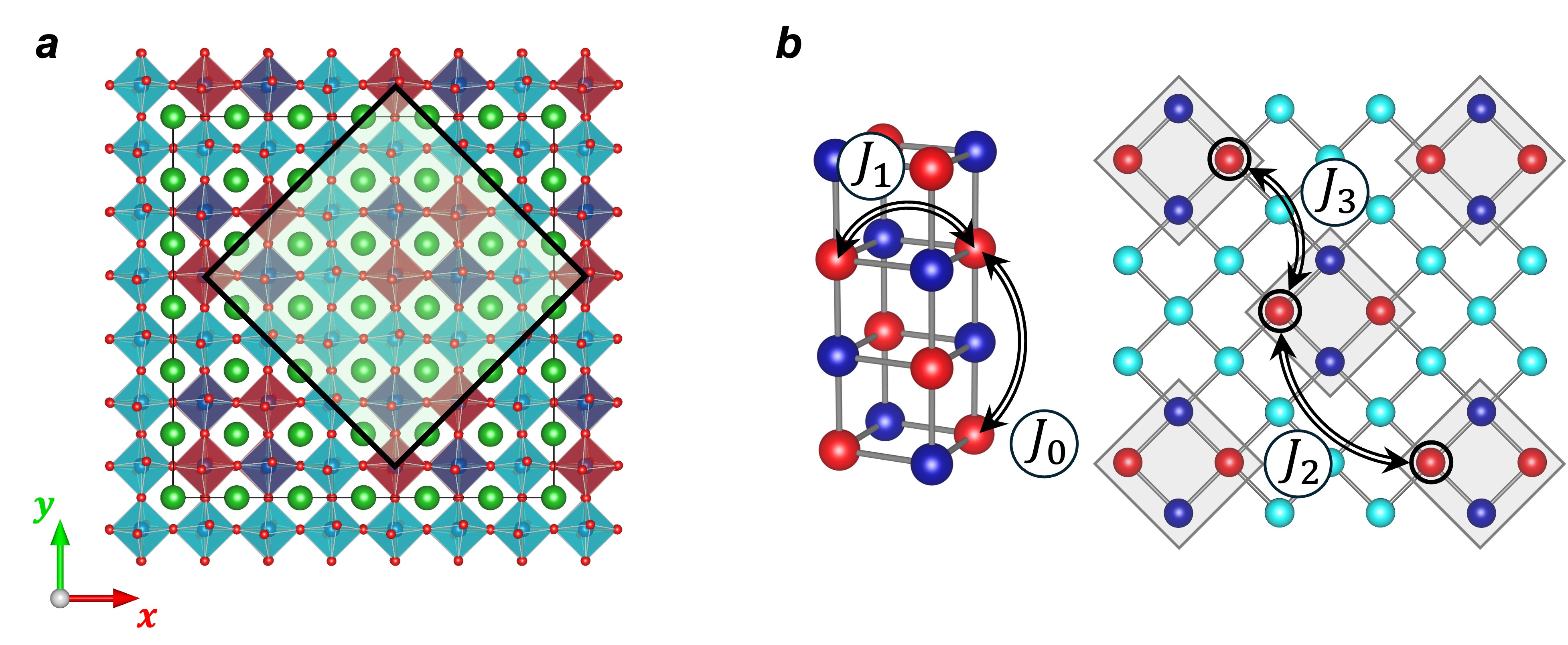}
    \caption{
    Constructed supercell derived from the $3\times3\times1$ ferromagnetic columnar model and the corresponding magnetic superexchange interactions. (a) The supercell is generated by halving the in-plane dimensions and doubling the out-of-plane dimension of the ground state configuration, allowing explicit treatment of out-of-plane exchange pathways. (b) Schematic illustration of the four distinct superexchange interactions. The intra-column interactions include the short-range out-of-plane $180^\circ$ interaction ($J_0$) and the short-range $90^\circ$ interaction ($J_1$). The inter-column interactions consist of the longer-range $180^\circ$ interaction ($J_2$) and the L-shaped $90^\circ$ interaction ($J_3$).
    }
    \label{fig:SEC}
\end{figure}

\begin{table}[ht]
    \centering
    
    \begin{tabular}{c|c|c}
         &
        Configuration &
        Equation \\
    \hline
         &
        Ground State &
        $8J_0 + 40J_1+8J_2+40J_3$ \\
        
        1 &
        Flip 2 from one column &
        $4J_0 + 20J_1+4J_2+24J_3$ \\
        
        2 &
        Flip 3 from one column &
        $2J_0 + 22J_1+2J_2+10J_3$ \\
        
        3 &
        Flip 4 from each column &
        $-8J_0 + 8J_1+8J_2+8J_3$ \\

        4 &
        Flip 1 from one and 2 from the other &
        $6J_0 + 10J_1+2J_2+18J_3$ \\

        5 &
        Flip 1 from one and 5 from the other &
        $4J_0 + 10J_1-8J_3$ \\

        6 &
        Flip 2 from one and 3 from the other &
        $-2J_0 +6J_1+2J_2+6J_3$ \\

        7 &
        Flip 3 from one and 4 from the other  &
        $2J_0 -6J_1+2J_2+2J_3$ \\

    \end{tabular}
    \caption{Magnetic configurations considered for the extraction of superexchange coupling parameters and the corresponding energy equations. The coefficients $n_i$ in the energy expressions are obtained by subtracting the interaction counts of the ground state configuration from those of each respective magnetic configuration.
    }

    \label{tab:js}
\end{table}

\begin{figure}[!h]
    \centering
    \includegraphics[width=0.6\textwidth]{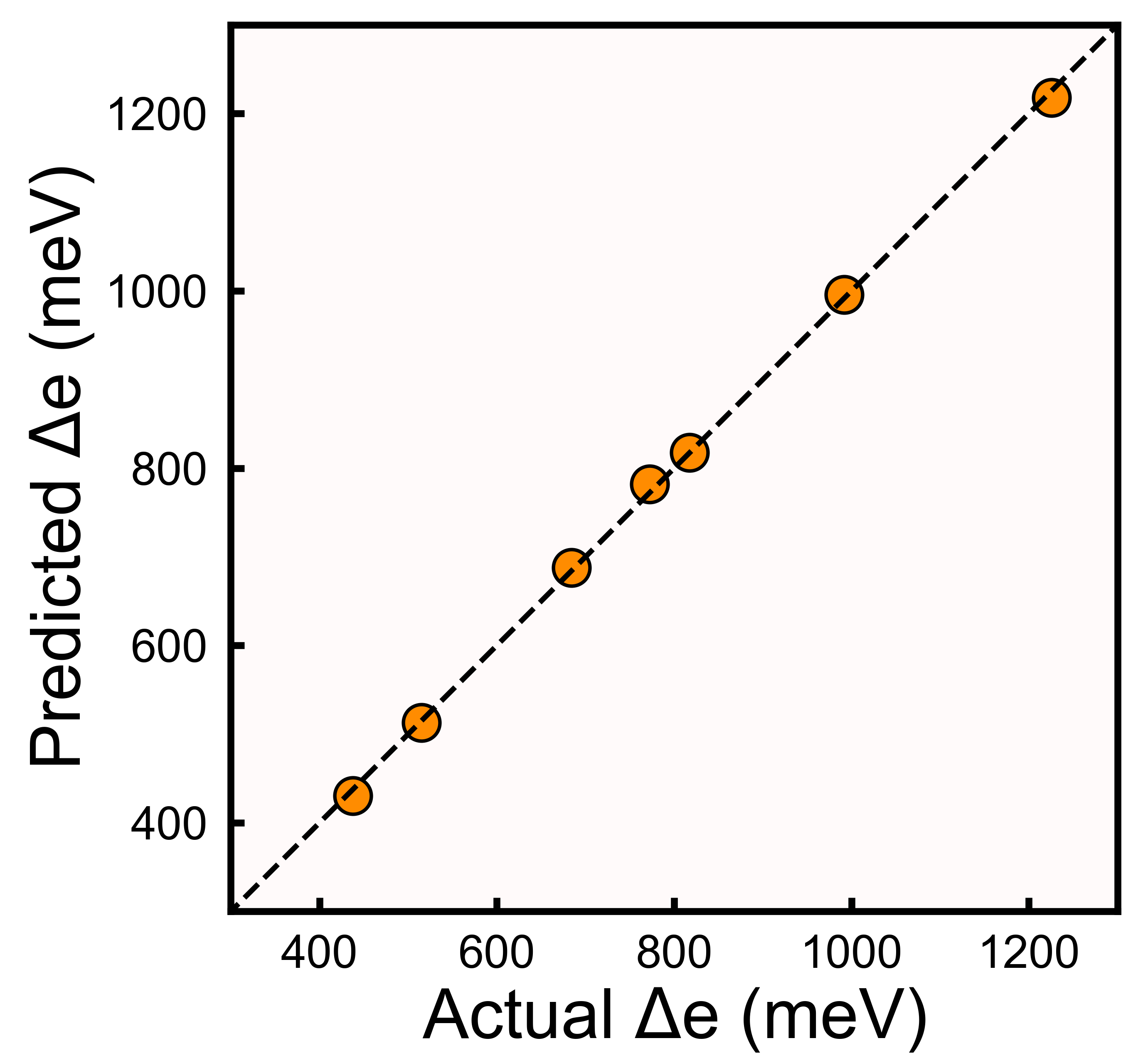}
    \caption{
    Accuracy of the least-squares fit used to extract the superexchange coupling parameters. The DFT-calculated energy differences for the seven magnetic configurations are plotted against the values predicted by the fitted Ising model. The excellent linear correlation, with a slope close to unity and negligible residuals, confirms the robustness and accuracy of the extracted exchange parameters with $R^2=0.9994$.
    }
    \label{fig:lsfit}
\end{figure}

\section{Superexchange Coupling Calculations}

To determine all superexchange coupling parameters, the original $3\times3\times1$ supercell is insufficient to capture the out-of-plane $180^\circ$ superexchange interaction ($J_0$) due to the imposed periodic boundary conditions along the $z$ direction and the resulting self-interaction of the HS Co ions. To overcome this limitation, a smaller but equivalent supercell was derived from the ground state configuration, as shown in \fig{SEC}(a). In this supercell, the in-plane dimensions were reduced by a factor of two, while the out-of-plane dimension was doubled, enabling explicit treatment of the out-of-plane $180^\circ$ superexchange pathway. In this supercell, two ferromagnetic columns are present.

As illustrated in \fig{SEC}(b), four distinct superexchange interactions are present in this geometry. To extract these coupling parameters, the spin orientation of selected HS Co ions was flipped to generate seven additional magnetic configurations relative to the ground state, yielding a total of seven independent energy equations. The corresponding spin configurations and equations are summarized in \tab{js}. The resulting overdetermined system of equations was solved using a least-squares fitting procedure, allowing all four superexchange coupling constants to be determined. The quality of the fit is demonstrated in \fig{lsfit}.

\FloatBarrier
\bibliographystyle{unsrtnat}

\end{document}